\documentclass[twocolumn,english,amssymb,aps,prl,superscriptaddress,tightenlines,notitlepage]{revtex4-2}
\usepackage{lmodern}
\usepackage[T1]{fontenc}
\usepackage[latin9]{inputenc}
\usepackage{color}
\usepackage{babel}
\usepackage{verbatim}
\usepackage{amsmath}
\usepackage{amssymb}
\usepackage{cancel}
\usepackage{graphicx}
\usepackage{esint}
\usepackage[unicode=true,pdfusetitle,
 bookmarks=true,bookmarksnumbered=true,bookmarksopen=false,
 breaklinks=false,pdfborder={0 0 1},backref=false,colorlinks=true]
 {hyperref}
\hypersetup{
 linkcolor=magenta,urlcolor=blue,citecolor=blue,pdfstartview={FitH},hyperfootnotes=false,bookmarks=False}

\makeatletter

\usepackage{babel}
\date{\today}
\usepackage{xcolor}
\usepackage{graphicx}
\allowdisplaybreaks 
\renewcommand{\big}{\bBigg@\@ne}
\renewcommand{\Big}{\bBigg@{1.5}}
\renewcommand{\bigg}{\bBigg@\tw@}
\renewcommand{\Bigg}{\bBigg@{2.5}}

\newcommand{\biggg}{\bBigg@\thr@@}
\newcommand{\Biggg}{\bBigg@{3.5}}

\makeatother

\begin{document}
\title{Inversion-asymmetric itinerant antiferromagnets by the space group symmetry}
\author{Changhee Lee}
\affiliation{Department of Physics and the MacDiarmid Institute for Advanced Materials
and Nanotechnology, University of Otago, P.O. Box 56, Dunedin 9054,
New Zealand}

\author{P. M. R. Brydon}
\affiliation{Department of Physics and the MacDiarmid Institute for Advanced Materials
and Nanotechnology, University of Otago, P.O. Box 56, Dunedin 9054,
New Zealand}
\begin{abstract}
We investigate the appearance of an inversion-asymmetric antiferromagnetism due to an itinerant mechanism in nonsymmorphic systems with magnetic ions at Wyckoff position of multiplicity 2. The key symmetries which underpin the existence of such phases are established, and we derive a Landau free energy from a general microscopic electronic Hamiltonian. Our analysis reveals that the stable antiferromagnetic order is largely determined by the symmetries of Wyckoff position, the nature of the nesting between electronic bands, and the presence of anisotropy or nesting in high-symmetry planes of the Brillouin zone. We illustrate our conclusions with specific microscopic models. 
\end{abstract}
\maketitle

\section{Introduction}

There is currently intense interest in magnetically-ordered states which lift the spin degeneracy of the electronic bands without possessing a net magnetization. This comes in two varieties, depending on the alignment of the microscopic moments. Collinear alignments correspond to the celebrated altermagnets~\citep{Hayami2019,Yuan2020,Mazin2021,Smejkal2022_Beyond,Smejkal2022_AnomalousHall}, where broken time-reversal symmetry produces large even-parity spin-splitting, which may be of great technological importance in spintronics applications~\citep{GonzalezHernandez2021,Karube2022,Smejkal2022_Magnetoresistance}. In the latter, a noncollinear magnetic order breaks inversion symmetry, and can consequently host odd-parity spin-splittings without the need for relativistic spin-orbit coupling~\citep{Hayami2020,hellenes2024pwavemagnets,Brekke2024,chakraborty2024highlyefficientnonrelativisticedelstein,Yue2025}. 
Compared to altermagnets, the study of these $p$-wave or inversion-asymmetric antiferromagnetic (IA-AFM) phases is much less advanced.

The IA-AFM must involve a combination of two magnetic orderings of opposite parity, which prevents the construction of an effective inversion symmetry by combining the normal-state inversion with a translation. As pointed out in Ref.~\citep{Yue2025}, IA-AFM can arise without fine-tuning in nonsymmorphic crystals, as many unit-cell-doubling magnetic orders belong to irreducible representations (irreps) which contain basis states of opposite parity. They proposed magnetic ions at Wyckoff positions of multiplicity two of these space groups as a minimal model for an IA-AFM, classified the corresponding spin-splittings induced by different unit-cell-doubling ordering vectors, and identified candidate materials.

Although nonsymmorphic systems are promising for the realization of IA-AFM, the mechanism by which these phases are stabilized remains unclear. Indeed, a pressing issue for the field is the absence of simple microscopic models that can provide such insight. Thus far, proposed models for IA-AFM involve conduction electrons coupled via exchange interactions to an odd-parity ordering of localized moments~\citep{Brekke2024,Nagae2025}. While this picture may be appropriate for the candidate material CeNiAsO~\cite{chakraborty2024highlyefficientnonrelativisticedelstein,Luo2011}, it is less suited to iron-based materials~\citep{Yue2025,Cvetkovic2013,Stadel2022}, where the magnetism has a more itinerant character. The itinerant scenario is intriguing, as the same electrons which display the odd-parity spin-splitting also directly generate the magnetic order while being subject to the symmetry of crystal. Here we propose a minimal microscopic theory for IA-AFM to be a tight-binding model of Wyckoff positions of multiplicity two with an on-site Hubbard interaction. In contrast to the localized-spin picture, where the interactions which stabilize the IA-AFM order are complicated and challenging to derive~\citep{Eszter2020,Szilva2023,Hatanaka2024Arxiv}, in our model the conditions for the IA-AFM can be directly related to the normal-state electronic structure and are thus significantly influenced by the symmetries of the underlying space group.

In this work, we use symmetry considerations to develop a minimal microscopic theory for the emergence
of the inversion-broken antiferromagnetic states via itinerant mechanism,
and uncover the conditions favourable to IA-AFM states. In Sec.~\ref{sec:Phenomenological-model}, we
identify the group-theoretical condition which ensures the existence of space
group irreps involving two components
with opposite parities~\citep{Szabo2023,Venderbos2016,Cvetkovic2013,Serbyn2013}. We hence show that the antiferromagnetic
states induced from such a mixed-parity irrep can be described by the phenomenological
Landau free energy proposed in Ref.~\citep{Yue2025}. We re-derive this free energy in Sec.~\ref{sec:Microscopic-model} starting from a microscopic electronic Hamiltonian for a two-sublattice system with Hubbard interaction. By a symmetry analysis of the normal-state hopping parameters and consideration of the various types of nesting instabilities available in this system, we identify conditions for the appearance of each of three antiferromagnetic states. We illustrate our analysis with two examples, exemplifying different behaviour of the sublattices under inversion. 

\section{Phenomenological model}

\label{sec:Phenomenological-model}

\begin{figure}
\includegraphics[width=0.4\textwidth]{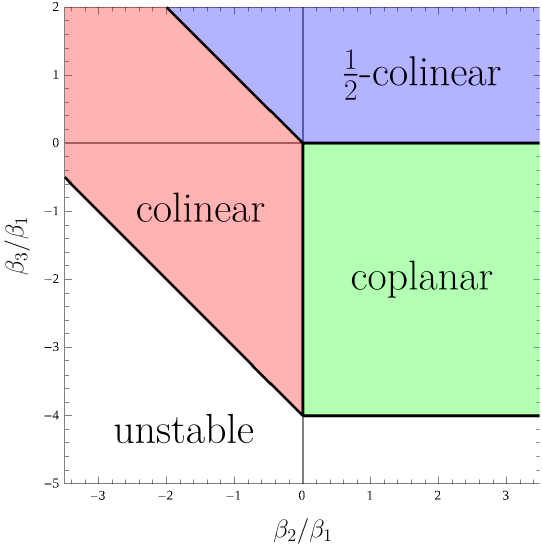} \caption{\label{fig:PhaseDiagram}Phase diagram of the Landau free energy in
Eq.~\eqref{eq:LandauFree};  $\beta_{1}>0$ is assumed. The coloured
regions correspond to the states enumerated in the text.}
\end{figure}

In centrosymmetric nonsymmorphic space groups, there are two-component
irreps for order parameters with unit-cell doubling ordering vector
$\vec{Q}$ which include both even- and odd-parity components~\citep{Szabo2023,Venderbos2016,Cvetkovic2013,Serbyn2013}.
These mixed-parity irreps occur when there is a symmetry $\{C|\vec{\tau}\}$, where $\vec{\tau}$ is a half-lattice translation, which leaves $\vec{Q}$ satisfying
$2\vec{Q}\cdot\vec{\tau}\in(2\mathbb{Z}+1)\pi$,
invariant up to a reciprocal lattice vector. Then, taking the center
of the inversion ${\cal I}$ as the origin, a group theoretical identity
\begin{equation}
\{{\cal I}|0\}\{C|\vec{\tau}\}=\{e|-2\vec{\tau}\}\{C|\vec{\tau}\}\{{\cal I}|0\}\label{eq:SymmetryCommutaion}
\end{equation}
with an identity operator $e$ guarantees that every irrep at momentum
$\vec{Q}$ involves both even and odd parity basis functions.

To show this, we first note that since nonsymmorphic symmetries do not leave a point invariant, at least two sublattices must be present in a unit cell. Furthermore, Eq.~\eqref{eq:SymmetryCommutaion} dictates that $\{{\cal I}|0\}$
and $\{C|\vec{\tau}\}$ cannot simultaneously be members of the site
symmetry group of a sublattice. This allows us to classify the
relevant Wyckoff positions (WP) of multiplicity two as either type-1
if inversion transposes the sublattices, or type-2 otherwise. In the following we specialize to the type-1 case; the argument for the type-2 case is presented in Appendices~\ref{App:MixedRep} and \ref{App:OddPower_SA_SB}. We let $f_{A}(\vec{r})$ be a real function defined on the $A$ sublattice only, oscillating with wavevector $\vec{Q}$ and invariant under a site-symmetry $\{C|\vec{\tau}\}$ of the sublattice
$A$; the function $f_{B}(\vec{r})\equiv\{{\cal I}|0\}f_{A}(\vec{r})$
is the analogue on the $B$ sublattice. Then, the even-parity function $f(\vec{r})=f_{A}(\vec{r})+f_{B}(\vec{r})$ can represent the density of an order
parameter with ordering vector $\vec{Q}$ defined over the entire
lattice. As $\vec{Q}$ is invariant under $C$, $\{C|\vec{\tau}\}f(\vec{r})$
also describes an order parameter density with ordering vector $\vec{Q}$,
whose parity is given by 
\begin{equation}
\{{\cal I}|0\}\{C|\vec{\tau}\}f(\vec{r})=e^{2i\vec{Q}\cdot\vec{\tau}}\{C|\vec{\tau}\}f(\vec{r})=-\{C|\vec{\tau}\}f(\vec{r}),
\end{equation}
where we use Eq.~\eqref{eq:SymmetryCommutaion}, the even parity of $f(\vec{r})$, and the condition on $\vec{Q}$. Since $\{C|\vec{\tau}\}f(\vec{r})$ has odd parity, the irrep containing both $f(\vec{r})$ and $\{C|\vec{\tau}\}f(\vec{r})$ does not have a definite parity. 

Given the basis for a space group irrep, an antiferromagnetic array of local moment can be expressed as $\vec{S}_{A}f_{A}(\vec{r})+\vec{S}_{B}f_{B}(\vec{r})$
where $\vec{S}_{A}$ and $\vec{S}_{B}$, serving as the order parameters,
represent the magnetic moments on sublattice $A$ and $B$, respectively.
When the spin-rotation symmetry at the microscopic level is present~\citep{Smejkal2022,Smejkal2022_Beyond}, we can relate the transformation rules of basis functions $f_{A,B}(\vec{r})$ under symmetry operations with those for $\vec{S}_{A,B}$ to obtain:
\begin{align}
\{{\cal I}|0\}\left(\begin{matrix}\vec{S}_{A}\\
\vec{S}_{B}
\end{matrix}\right)= & \left(\begin{matrix} 0 & 1\\
1 & 0
\end{matrix}\right)\left(\begin{matrix}\vec{S}_{A}\\
\vec{S}_{B}
\end{matrix}\right),\\
\{C|\vec{\tau}\}\left(\begin{matrix}\vec{S}_{A}\\
\vec{S}_{B}
\end{matrix}\right)= & \left(\begin{matrix}1 & 0\\
 0 & -1
\end{matrix}\right)\left(\begin{matrix}\vec{S}_{A}\\
\vec{S}_{B}
\end{matrix}\right).
\end{align}
These relations directly inform the Landau expansion of the free energy in terms of $\vec{S}_A$ and $\vec{S}_B$; in particular, 
terms involving odd powers of $\vec{S}_{A}\cdot\vec{S}_{B}$
or $|\vec{S}_{A}|^{2}-|\vec{S}_{B}|^{2}$ are odd under $\{{\cal I}|0\}$
or $\{C|\vec{\tau}\}$, respectively, and are thus forbidden. To quartic order, the Landau free energy is hence~\citep{Yue2025}:
\begin{align}
{\cal F}= & \alpha(T)(|\vec{S}_{A}|^{2}+|\vec{S}_{B}|^{2})+\beta_{1}(|\vec{S}_{A}|^{2}+|\vec{S}_{B}|^{2})^{2}\nonumber \\
 & +\beta_{2}(\vec{S}_{A}\cdot\vec{S}_{B})^{2}+\beta_{3}|\vec{S}_{A}|^{2}|\vec{S}_{B}|^{2}.\label{eq:LandauFree}
\end{align}
This expression also holds for the type-2 case. The validity of the quartic expansion Eq.~\eqref{eq:LandauFree} requires the stability conditions $\beta_{1}>0$, $\beta_{3}>-4\beta_{1}$, and $\beta_{2}+\beta_{3}>-4\beta_{1}$.

 The phase diagram shown in Fig.~\ref{fig:PhaseDiagram} exhibits three types of saddle point solutions: (i) the colinear phase with $\vec{S}_{A} = \pm\vec{S}_{B}$ for $\beta_{2}<0$, and $-4\beta_1<\beta_2+\beta_3<0$; (ii)  the half-colinear phase with either vanishing $\vec{S}_{A}$ or $\vec{S}_B$ when $\beta_3>0$ and $0<\beta_{3}+\beta_{2}$; (iii) the coplanar phase with $|\vec{S}_A|=|\vec{S}_B|$ and $\vec{S}_A\cdot\vec{S}_B=0$ for $0<\beta_{2}$ and $-4<\beta_{3}<0$. The spin-degeneracy of the electronic bands in the colinear states is ensured by the invariance under spin rotations about the axis of the staggered moment, as well as two-fold rotation about a perpendicular axis followed by a lattice translation~\citep{Sato2016}. Only the latter symmetry is available in the coplanar state, however, and so the spin degeneracy of the electronic bands can be lifted~\citep{hellenes2024pwavemagnets,Yue2025}. We note that a formally-similiar free energy describes the spin-density wave phases in the iron-pnictide superconductors~\citep{Lorenzana2008,Eremin2010,Giovannetti2011,Brydon2011,Wang2015}, where the colinear, half-colinear, and coplanar states are known as the magnetic stripe, spin-charge order, and orthomagnetic states, respectively~\citep{Lorenzana2008}. 

\section{Microscopic model}
\label{sec:Microscopic-model}
\subsection{General consideration}
To derive the free energy in Eq.~\eqref{eq:LandauFree}
from a microscopic electronic Hamiltonian, we consider a two-sublattice system containing all symmetry-allowed spin-independent hoppings and an on-site Hubbard interaction of strength $U$. We decouple
the interaction via a Hubbard-Stratonovich transformation by introducing the auxiliary bosonic fields $\vec{S}_{A}$ and $\vec{S}_{B}$, which correspond to the magnetic order parameters with ordering vector $\vec{Q}$. The resulting saddle-point Hamiltonian is 
\begin{equation}
\hat{H}  =\sum_{\vec{k}}\hat{\Psi}_{\vec{k}}^{\dagger}\left(\begin{matrix}h_{\vec{k}} & \Sigma\\
\Sigma^\dagger & h_{\vec{k}+\vec{Q}}
\end{matrix}\right)\hat{\Psi}_{\vec{k}}+\frac{|\vec{S}_{A}|^{2}+|\vec{S}_{B}|^{2}}{U}\label{eq:ElectronHam}
\end{equation}
where
\begin{align}
h_{\vec{k}} & =[\varepsilon_{\vec{k},0}\tau_{0}+t_{\vec{k},x}\tau_{x}+t_{\vec{k},y}\tau_{y}+\varepsilon_{\vec{k},z}\tau_{z}]\sigma_{0}, \label{eq:hk}\\
\Sigma & =\frac{1}{2}\vec{S}_{A}\cdot\vec{\sigma}(\tau_{0}+\tau_{z})+\frac{1}{2}\vec{S}_{B}\cdot\vec{\sigma}(\tau_{0}-\tau_{z}),
\end{align}
with $\hat{\Psi}_{\vec{k}}=(\hat{C}_{\vec{k}},\hat{C}_{\vec{k}+\vec{Q}})^{T}$
where $\hat{C}_{\vec{k}}=(\hat{C}_{\vec{k},A,\uparrow},\hat{C}_{\vec{k},A,\downarrow},\hat{C}_{\vec{k},B,\uparrow},\hat{C}_{\vec{k},B,\downarrow})^{T}$ is a spinor of electron annihilation operators.
$\tau_{i}$ and $\sigma_{i}$ are the Pauli matrices for the sublattice
and spin degrees of freedom, respectively. $\varepsilon_{\vec{k},0}$
and $\varepsilon_{\vec{k},z}$ represent the average and the difference
of the intra-sublattice hoppings (including the chemical
potential), while $t_{\vec{k},x}$ and $t_{\vec{k},y}$ originate
from the  inter-sublattice hoppings. The sublattice-nontrivial terms are particularly influenced by the space-group symmetry, as discussed in Appendix~\ref{App:sym_t}. The normal-state Hamiltonian describes a two-band system with dispersions $\xi_{\vec{k},\pm}=\varepsilon_{\vec{k},0}\pm\sqrt{t_{\vec{k},x}^{2}+t_{\vec{k},y}^{2} + \varepsilon_{\vec{k},z}^{2}}$. In the following we work in a gauge for the Bloch basis functions such that $h_{\vec{k}}$ is invariant under $\vec{k}\rightarrow\vec{k}+\vec{G}$ for any reciprocal lattice vector $\vec{G}$. This implies that the order parameters $\vec{S}_{A}$ and $\vec{S}_{B}$ are real-valued. We examine the alternative gauge choice used by Ref.~\citep{Yue2025} in Appendix~\ref{App:sym_t}.

Integrating out the fermionic fields and expanding the resulting free energy up to quartic
order in the order parameter, we obtain the general Landau free energy 
\begin{align}
{\cal F}= & \alpha_A|\vec{S}_A|^2 + \alpha_B|\vec{S}_B|^2 +\alpha_{2}\vec{S}_{A}\cdot\vec{S}_{B} \nonumber \\
 & +\beta_{1}(|\vec{S}_{A}|^{2}+|\vec{S}_{B}|^{2})^{2}+\beta_{2}(\vec{S}_{A}\cdot\vec{S}_{B})^{2}+\beta_{3}|\vec{S}_{A}|^{2}|\vec{S}_{B}|^{2}\nonumber\\
 &+\beta_{4}(|\vec{S}_{A}|^{4}-|\vec{S}_{B}|^{4})+\beta_{5}(|\vec{S}_{A}|^{2}+|\vec{S}_{B}|^{2})\vec{S}_{A}\cdot\vec{S}_{B}\nonumber \\
 & +\beta_{6}(|\vec{S}_{A}|^{2}-|\vec{S}_{B}|^{2})\vec{S}_{A}\cdot\vec{S}_{B}\,.\label{eq:generalF}
\end{align}
Explicit expressions for all these coefficients are given in Appendix~\ref{App:beta}.
Although Eq.~\eqref{eq:generalF} contains a number of terms which are absent from Eq.~\eqref{eq:LandauFree}, the symmetry of the sublattice nontrivial terms in $h_{\vec{k}}$ under $\{{\cal I}|0\}$
and $\{C|\vec{\tau}\}$ directly implies that the coefficients $\alpha_2$ and $\beta_{4,5,6}$ must vanish, and $\alpha_A=\alpha_B = U^{-1}-\chi$ with noninteracting spin susceptibility $\chi$. We hence recover the phenomenological free energy Eq.~\eqref{eq:LandauFree}, but the coefficients are now derived from a microscopic model. 

It is convenient to consider the following combinations of coefficients
\begin{align}
\chi= & \sum_{\vec{k}}\sum_{a,b=\pm}\frac{\tanh\frac{\xi_{\vec{k},a}}{2T}-\tanh\frac{\xi_{\vec{k}+\vec{Q},b}}{2T}}{\xi_{\vec{k},a}-\xi_{\vec{k}+\vec{Q},b}},\label{eq:chi}\\
\beta_{2}= & 2T\sum_{n,\vec{k}}\frac{|\vec{t}_{\vec{k}}\cdot\vec{t}_{\vec{k}+\vec{Q}}|^{2}-|\vec{t}_{\vec{k}}\times\vec{t}_{\vec{k}+\vec{Q}}|^{2}}{g(i\omega_n,\vec{k})},\label{eq:beta2}\\
\beta_{3}= & -2\gamma+2T\sum_{n,\vec{k}}\frac{|\vec{t}_{\vec{k}}\times\vec{t}_{\vec{k}+\vec{Q}}|^{2}}{g(i\omega_n,\vec{k})},\label{eq:beta3-beta1}\\
\beta_{2}+\beta_{3}= & -2\gamma+2T\sum_{n,\vec{k}}\frac{|\vec{t}_{\vec{k}}\cdot\vec{t}_{\vec{k}+\vec{Q}}|^{2}}{g(i\omega_n,\vec{k})},\label{eq:beta2+beta3-beta1}
\end{align}
where
\begin{equation}
\gamma=\sum_{\vec{k}}\sum_{a,b=\pm}\frac{-ab}{8|\vec{t}_{\vec{k}}||\vec{t}_{\vec{k}+\vec{Q}}|}\frac{\tanh\frac{\xi_{\vec{k},a}}{2T}-\tanh\frac{\xi_{\vec{k}+\vec{Q},b}}{2T}}{\xi_{\vec{k},a}-\xi_{\vec{k}+\vec{Q},b}},\label{eq:gamma}
\end{equation}
and  $T$ denoting temperature, $g(i\omega_n,\vec{k})=\prod_{\lambda=\pm}(i\omega_{n}-\xi_{\vec{k},\lambda})^{2}(i\omega_{n}+\xi_{\vec{k}+\vec{Q},\lambda})^{2}$, and $\vec{t}_{\vec{k}}=t_{\vec{k},x}\hat{x} + t_{\vec{k},y}\hat{y}$.

A second-order transition into the antiferromagnetic state occurs when $\chi = U^{-1}$; at weak-coupling, this corresponds to fillings where there is good nesting between the bands $\xi_{\vec{k},a}$ and $\xi_{\vec{k}+\vec{Q},b}$. The nesting can be classified as either interband ($a\neq b$) or intraband ($a=b$). Since a nesting of one band does not imply nesting of the other, intraband nesting generally only produces a divergence of one term in the sum over the bands in Eq.~\eqref{eq:chi}; in contrast, since interband nesting involves both bands, two terms in the sum diverge. We therefore expect that interband nesting is more favourable for realizing an antiferromagnetic state than intraband nesting. 

Assuming that the stability conditions are satisfied, the sign of Eqs.~\eqref{eq:beta2}-\eqref{eq:beta2+beta3-beta1} determine the stable spin configuration, as can be readily checked by comparing to Fig.~\ref{fig:PhaseDiagram}. These signs are partially determined by the character of the nesting through the quantity $\gamma$, which should dominate  Eqs.~\eqref{eq:beta3-beta1} and~\eqref{eq:beta2+beta3-beta1} at sufficiently high temperature: it is positive (negative) if interband (intraband) nesting occurs, and a positive (negative) $\gamma$ implies that the coplanar or colinear (half-colinear) phase is realized.

Near the optimal filling for interband nesting, the sign of $\beta_2$ decides between the colinear and coplanar phases. When the nesting is close to ideal (i.e. $\xi_{\vec{k},a} = -\xi_{\vec{k}+\vec{Q},-a}$ near the Fermi level), we can approximate $\beta_2$ as
\begin{equation}
\beta_{2} \approx2T\sum_{k}\frac{|\vec{t}_{\vec{k}}\cdot\vec{t}_{\vec{k}+\vec{Q}}|^{2}-|\vec{t}_{\vec{k}}\times\vec{t}_{\vec{k}+\vec{Q}}|^{2}}{(\omega_{n}^{2}+\xi_{\vec{k},+}^{2})^{2}(\omega_{n}^{2}+\xi_{\vec{k},-}^{2})^{2}}.\label{eq:simple_beta2}
\end{equation}
This reveals that the sign of $\beta_{2}$, and thus the stable antiferromagnetic state, is largely determined by
the sign of $|\vec{t}_{\vec{k}}\cdot\vec{t}_{\vec{k}+\vec{Q}}|^{2}-|\vec{t}_{\vec{k}}\times\vec{t}_{\vec{k}+\vec{Q}}|^{2}$. By consideration of the symmetries of the intersublattice terms (see Appendix~\ref{App:sym_t}), we find that $\vec{t}_{\vec{k}}\cdot\vec{t}_{\vec{k}+\vec{Q}}$
is even (odd) under inversion in type-1 (type-2) theories, while
it is even under the time-reversal in both cases. As a result, $\vec{t}_{\vec{k}}\cdot\vec{t}_{\vec{k}+\vec{Q}}=0$
in the type-2 case, which implies that $\beta_{2}$ is negative, and so the coplanar phase is disfavoured in such systems at the optimal doping for interband nesting.

\begin{figure}
\includegraphics[width=0.98\columnwidth]{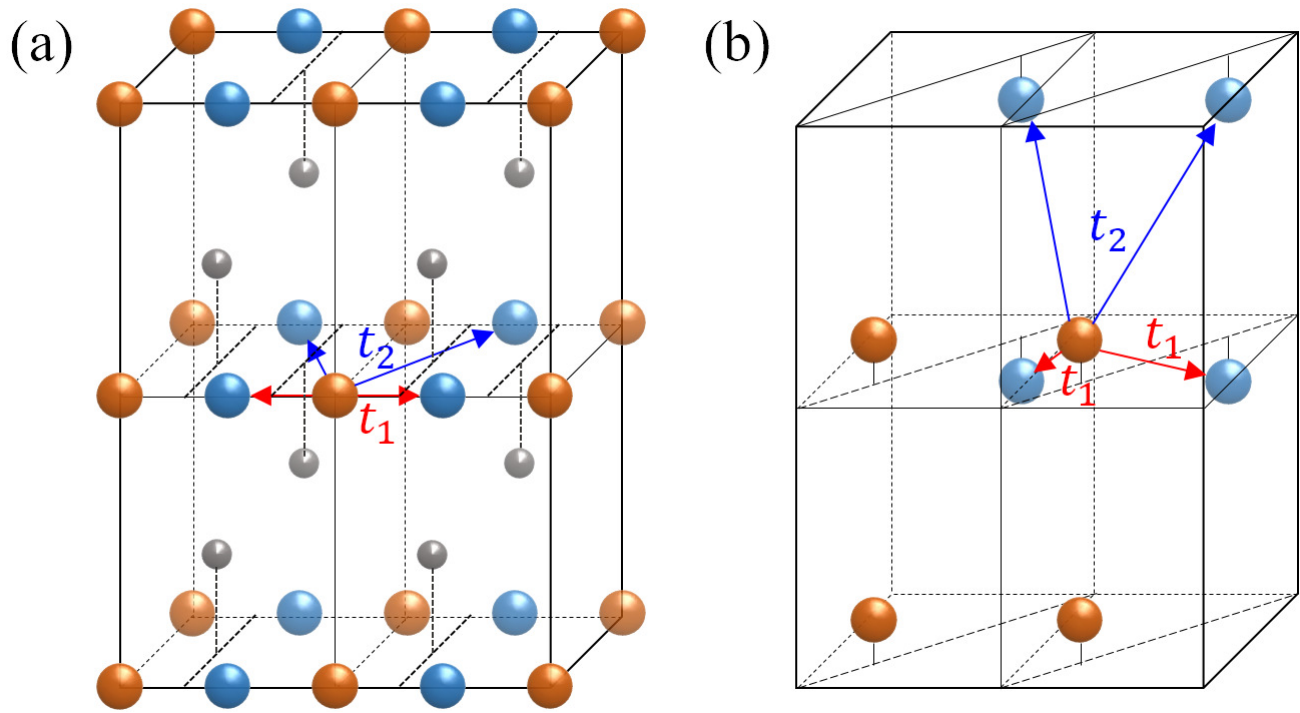}\caption{\label{fig:Example}(a) Hopping parameters $t_{1}$ and $t_{2}$ in
Eq. \eqref{eq:hopping_SG51} of the tight-binding model of space group
51, (b) Hopping parameters $t_{1}$ and $t_{2}$ in Eq. \eqref{eq:hopping_SG129}
of the tight-binding model of space group 129. To make the nonsymmorphic
structure in (a) more evident, additional atoms depicted
by gray balls are placed at the Wyckoff position 2f.}
\end{figure}

Focusing now on a lattice with type-1 WP, it is interesting to note that although the coplanar phase is likely to emerge when $|\vec{t}_{\vec{k}}\times\vec{t}_{\vec{k}+\vec{Q}}|$ is small compared to $|\vec{t}_{\vec{k}}\cdot\vec{t}_{\vec{k}+\vec{Q}}|$, the spin-splitting of the bands due to the magnetic
order is proportional to $|\vec{t}_{\vec{k}}\times\vec{t}_{\vec{k}+\vec{Q}}||\vec{S}_{A}\times\vec{S}_{B}|$. Thus, the spin-splitting of the bands is in tension with the stability of the coplanar phase.
To this apparent conflict, we identify two possible solutions: (i) The hopping parameters are such that $|\vec{t}_{\vec{k}}\cdot\vec{t}_{\vec{k}+\vec{Q}}|^{2}>|\vec{t}_{\vec{k}}\times\vec{t}_{\vec{k}+\vec{Q}}|^{2}$
is achieved for general momentum $\vec{k}$; (ii) Symmetry enforces $|\vec{t}_{\vec{k}}\times\vec{t}_{\vec{k}+\vec{Q}}|$
to vanish close to the regions where the nesting is significant. 
We find that the first solution is especially favoured when there is a significant anisotropy in the hopping, which can be easily realized in orthorhombic or tetragonal lattices.

The second solution is applicable if there is a site-preserving symmetry
$\{R|\vec{\tau}_{R}\}$ leaving $\vec{Q}$ invariant up to a reciprocal lattice vector and $2\vec{Q}\cdot\vec{\tau}_R\in2\pi\mathbb{Z}$; such a symmetry is always present in primitive tetragonal and cubic systems, as shown in Appendix~\ref{App:existence_twofold_sym}.
Combined with the inversion or the time-reversal symmetry, it forces
$\vec{t}_{\vec{k}}\times\vec{t}_{\vec{k}+\vec{Q}}$ to vanish on $\vec{k}$
which is invariant under ${\cal I}R$. Thus, the coplanar phase can
be more stable than the colinear phase if these planes lie close to the regions of optimal interband nesting.

\subsection{Example: Space Group 51}

We first consider a microscopic model consistent with the orthorhombic space group 51. Both types of WPs are available in this space group, but here we specialize to the type-2 position 2a. A possible realization and intersublattice hopping integrals is shown in Fig.~\ref{fig:Example}(a); for simplicity we assume negligible hopping along the $c$-axis, which can be realized for $c\gg a$, $b$ with three lattice constants $a$, $b$, and $c$. The intersublatice terms in Eq.~\eqref{eq:hk} are
\begin{equation}
t_{\vec{k},x}-it_{\vec{k},y}=  [t_{1}+t_{2}\cos (k_{y}b)](1+e^{-ik_{x}a}),\label{eq:hopping_SG51}
\end{equation}
The lowest-order intra-sublattice hopping term  $\varepsilon_{\vec{k},z}$ is $\sim\sin(k_{x}a)\sin(k_{z}c)$ and we thus ignore it in the following. 

Values of $\vec{Q}$ consistent with the validity of Eq.~\eqref{eq:LandauFree} are $\vec{Q}=(\frac{\pi}{a},m\frac{\pi}{b},n\frac{\pi}{c})$ with $m$, $n=0,1$. It can be explicitly verified that $\vec{t}_{\vec{k}}\cdot\vec{t}_{\vec{k}+\vec{Q}}=0$ for these ordering vectors, as required by time-reversal and inversion symmetry.
Fig. \ref{fig:chi_and_beta2}(a) displays $\chi$, $\beta_{2}$,
and $\gamma$ for $\vec{Q}=(\frac{\pi}{a},\frac{\pi}{b},0)$ as a function of the chemical potential $\mu$ for a specific two-dimensional tight-binding model. The maximum in $\chi$ at $\mu=0$ corresponds to interband nesting, while the two shoulder features arise from intraband nesting, consistent with our expectation that interband nesting favours antiferromagnetic order. As also expected for type-2 WPs, the interband nesting at optimal doping gives rise to colinear magnetic order; half-colinear order dominates around the shoulder features, again consistent with our expectation for dominant intraband nesting. Thin slivers of coplanar order are visible at the edges of the half-colinear phase; although the numerator of $\beta_2$ is strictly negative for type-2 WPs, the approximation Eq.~\eqref{eq:simple_beta2} is not available near intraband nesting instabilities, and so it is possible that $\beta_2$ can have a positive value here.

\subsection{Example: Space Group 129}

The tetragonal space group 129 has only type-1 WPs, of which here we consider WP 2c. The hoppings shown in Fig.~\ref{fig:Example}(b) generate the intersublatice terms
\begin{equation}
t_{\vec{k},x}-it_{\vec{k},y}=  (t_{1}+t_{2}e^{-ik_{z}c})\prod_{i=x,y}(1+e^{-ik_{i}a}),\label{eq:hopping_SG129}
\end{equation}
Here we again consider a model with significant $c$-axis anisotropy, so that only the intersublattice terms have $k_z$-dependence. This model is directly relevant to the Ce ions in CeRh$_2$As$_2$; nuclear magnetic resonance and muon spin resonance studies report signature of antiferromagnetism in this material  below $T\sim 0.5\rm{K}$ which is much lower than the Kondo temperature~$\sim 30\rm{K}$ ~\citep{Khim2024uSR,Ogata2023,Kibune2022}. Furthermore, a recent inelastic neutron scattering measurements suggests that $\vec{Q}=(\frac{\pi}{a},\frac{\pi}{a},0)$ is the ordering vector of the magnetic phase~\citep{TChen2024Expt}.

Eq.~\eqref{eq:LandauFree} holds for $\vec{Q}=(\frac{\pi}{a},\frac{\pi}{a},n\frac{\pi}{c})$ with $n=0,1$. For $\vec{Q}=(\frac{\pi}{a},\frac{\pi}{a},0)$ we find that $\vec{t}_{\vec{k}}\times\vec{t}_{\vec{k}+\vec{Q}}=0$, and so the coplanar phase is favoured but does not produce a spin-splitting of the electronic dispersion. This is an artifact of our choice of intersublattice hoppings Eq.~\eqref{eq:hopping_SG129}. Including longer-range hopping, the cross product is nonzero at general points
in the Brillouin zone; since longer-range hopping should be small, we expect that the coplanar phase is stable and accompanied by spin-splitting. Moreover, the two-fold symmetry $\{C_{2z}|\frac{1}{2}\frac{1}{2}0\}$
still enforces $|\vec{t}_{\vec{k}}\times\vec{t}_{\vec{k}+\vec{Q}}|$
to vanish for $k_z=0$ or $\frac{\pi}{c}$ planes. Consequently, the coplanar phase will remain strongly favoured if the best interband nesting condition occurs in basal planes.

\begin{figure}
\includegraphics[width=0.98\columnwidth]{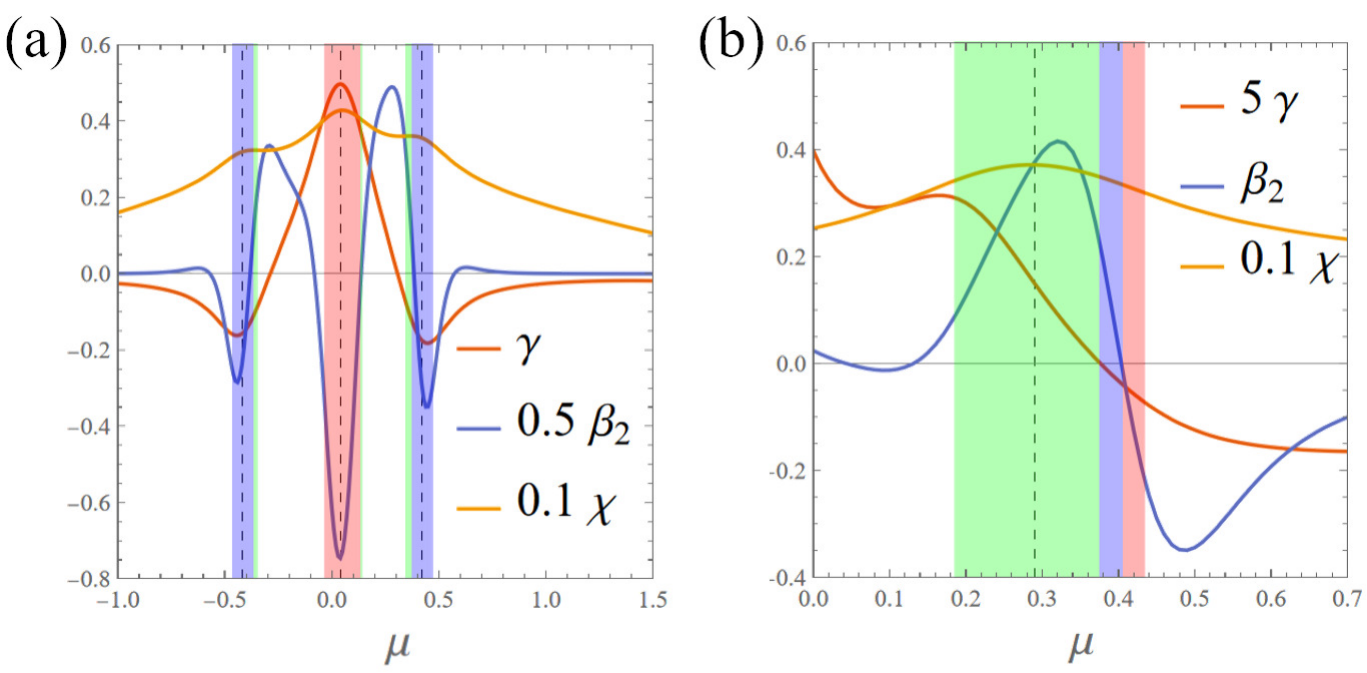}\caption{\label{fig:chi_and_beta2} $\chi$, $\beta_{2}$, and $\gamma$ at
$T=0.05t$ with respect to the chemical potential $\mu$ in Eq.~\eqref{eq:ElectronHam}. (a) is obtained from a model of type-2 WP in space group 51 with $\vec{Q}=(\frac{\pi}{a},\frac{\pi}{b},0)$. $\varepsilon_{\vec{k},0}=2t(\cos k_{x}a+0.1\cos k_{y}b)-\mu$
and $(t_{1},t_{2})=(0.3,0.05)t$ is used. 
(b) shows the result from
a system of type-1 WP in space group 129 with $\vec{Q}=(\frac{\pi}{a},\frac{\pi}{a},0)$. $\varepsilon_{\vec{k},0}=2t(\cos k_{x}a+\cos k_{y}a)-\mu+4t'\cos k_{x}a\cos k_{ya}$
and $(t',t_{1},t_{2})=(0.05,0.1,0.5)t$ is used.
 The red, green, and blue shading represent the states with the same shading in Fig.~\ref{fig:PhaseDiagram}. The stability conditions for the free energy Eq.~\eqref{eq:LandauFree} are violated in the unshaded region. Some of $\chi$, $\gamma$, and $\beta_{2}$ are rescaled
for enhanced visibility.}
\end{figure}

Fig.~\ref{fig:chi_and_beta2}(b) displays typical $\chi$, $\beta_2$, and $\gamma$ as a function of $\mu$ for the $\vec{Q}=(\frac{\pi}{a},\frac{\pi}{a},0)$ case. For this demonstration, we suppose that the intra-sublattice term $\varepsilon_{\vec{k},0}$ dominates Eq.~\eqref{eq:hk}. The system shows good interband nesting near the maximum of $\chi$, which combined with the vanishing of the cross produce $\vec{t}_{\vec{k}}\times\vec{t}_{\vec{k}+\vec{Q}}$, stabilizes the coplanar phase, and we accordingly find that the coplanar phase is realized near the maximum of $\chi$. Adjusting the chemical potential to create intraband nesting allows the other two phases appear.

A spin-splitting in the coplanar phase with ordering vector $\vec{Q}=(\frac{\pi}{a},\frac{\pi}{a},\frac{\pi}{c})$ appears already for the hopping in Eq.~\eqref{eq:hopping_SG129}; here we find that
\begin{align}
\vec{t}_{\vec{k}}\cdot\vec{t}_{\vec{k}+\vec{Q}} = & 4(t_{2}^{2}-t_{1}^{2})\sin (k_{x}a) \sin (k_{y}a) \\
\vec{t}_{\vec{k}}\times\vec{t}_{\vec{k}+\vec{Q}} = & 8t_{1}t_{2}\sin (k_{x}a) \sin (k_{y}a) \sin (k_{z}c)\hat{z}
\end{align}
Although a positive sign of the numerator of $\beta_2$ is not guaranteed, the same symmetry argument as above ensures this happens if the interband nesting occurs on the $k_z=0$ or $\frac{\pi}{c}$ planes. Moreover, the coplanar state is also favoured when there is strong $c$-axis anisotropy with $t_{2}/t_{1} \gg (\ll) 1$, which is plausible if the lattice constant $c$ is much larger than $a$. The $f$-wave spin-splitting of the electronic dispersion may be considerably larger than for the $\vec{Q}=(\frac{\pi}{a},\frac{\pi}{a},0)$ order, since it does not require longer-range hopping.

\section{Discussion}

In this work we have investigated the conditions required for the existence of commensurate IA-AFM states which can be reached in a single second-order transition from the normal state. We have clarified the symmetry conditions which allows this to occur, specifically the existence of mixed-parity irreps, and used them to construct minimal phenomenological and microscopic models in terms of magnetic ions occupying WP of multiplicity 2. Our general microscopic model, based on the itinerant mechanism, reveals an essential role for the intersublattice hopping terms, giving two  simple conditions that strongly favour an odd-parity coplanar state with spin-splitting of the electron bands: strong interband nesting, and $|\vec{t}_{\vec{k}}\cdot\vec{t}_{\vec{k}+\vec{Q}}|^2>|\vec{t}_{\vec{k}}\times\vec{t}_{\vec{k}+\vec{Q}}|^2$. This places significant constraints on the system, with the coplanar state being unfavourable when inversion is a site-symmetry, and also shows a tension between the nonrelativistic spin-splitting of the electronic dispersion and the stability of the coplanar state. We have illustrated and validated these conclusions by two specific models. Our work provides a general symmetry-informed guide for the realization of inversion-asymmetric antiferromagnetism with spin-split bands.

\acknowledgments
The authors thank Y. Yu, D. F. Agterberg, Brian M. Andersen, and Andreas Kreisel for stimulating discussions. This work was supported by the Marsden Fund Council from Government funding, managed by Royal Society Te Ap\={a}rangi, Contract No. UOO2222.

\bibliography{ref}

\clearpage{}

\appendix

\section{Existence of the mixed-parity irreducible representations}
\label{App:MixedRep}
\subsection{Sublattice-transposing inversion}
Here, we first show by direct construction that
the presence of the symmetries $\{{\cal I}|0\}$ and $\{C|\vec{\tau}\}$
enforces the existence of mixed-parity irreps for states with ordering
vector $\vec{Q}$ satisfying $\exp(2i\vec{Q}\cdot\vec{\tau})=-1$.
For this purpose, we consider two periodic functions $f_{A}$ and $f_{B}$ with propagation vector $\vec{Q}$, which
have values only at the two sublattices $A$ and $B$, respectively.
\begin{align}
f_{A}(\vec{r})= & \sum_{\vec{R}}\delta(\vec{r}-\vec{R}-\vec{r}_{A})f_{\vec{Q}}(\vec{R}),\label{eq:fA}\\
f_{B}(\vec{r})\equiv & \{{\cal I}|0\}f_{A}(\vec{r})=\sum_{\vec{R}}\delta(\vec{r}-\vec{R}-\vec{r}_{B})f_{\vec{Q}}(\vec{R}),\label{eq:fB}
\end{align}
where $\vec{R}$ denotes the lattice vectors and $f_{\vec{Q}}(\vec{R}+\vec{R}')=e^{i\vec{Q}\cdot\vec{R}'}f_{\vec{Q}}(\vec{R})$
for a lattice vector $\vec{R}'$. Here, we use $e^{i\vec{Q}\cdot\vec{R}}=e^{-i\vec{Q}\cdot\vec{R}}$
and choose an inversion center as the origin so that $\vec{r}_{A}=-\vec{r}_{B}+\vec{t}$
for some lattice vector $\vec{t}$. Then, the symmetric combination
of $f_{A}$ and $f_{B}$ results in an even-parity function.

By applying a site symmetry $\{C|\vec{\tau}\}$ of the site $\vec{r}_{A}$
to $f_{A}+f_{B}$ , we introduce a function $\tilde{f}_{A}(\vec{r})+\tilde{f}_{B}(\vec{r})$
with 
\begin{align}
\tilde{f}_{A}(\vec{r})\equiv & \{C|\vec{\tau}\}f_{A}(\vec{r}),\\
\tilde{f}_{B}(\vec{r})\equiv & \{C|\vec{\tau}\}f_{B}(\vec{r}).
\end{align}
The parity of $\tilde{f}_{A}(\vec{r})+\tilde{f}_{B}(\vec{r})$ is determined by using 
\begin{equation}
\{{\cal I}|0\}\{C|\vec{\tau}\}=\{C|\vec{\tau}\}\{{\cal I}|0\}\{e|2C^{-1}\vec{\tau}\},
\end{equation}
which yields \begin{widetext}
\begin{align}
\{{\cal I}|0\}\tilde{f}_{A}(\vec{r}) & =\{{\cal I}|0\}\{C|\vec{\tau}\}f_{A}(\vec{r})=e^{-i2(C\vec{Q}\cdot\vec{\tau})}\{C|\vec{\tau}\}f_{B}(\vec{r})=e^{-i2(C\vec{Q}\cdot\vec{\tau})}\tilde{f}_{B}(\vec{r}),\\
\{{\cal I}|0\}\tilde{f}_{B}(\vec{r}) & =\{{\cal I}|0\}\{C|\vec{\tau}\}f_{B}(\vec{r})=e^{-i2(C\vec{Q}\cdot\vec{\tau})}\{C|\vec{\tau}\}f_{A}(\vec{r})=e^{-i2(C\vec{Q}\cdot\vec{\tau})}\tilde{f}_{A}(\vec{r}).
\end{align}
\end{widetext}Since $e^{-2i(C\vec{Q}\cdot\vec{\tau})}=e^{-2i\vec{Q}\cdot\vec{\tau}}$
for a lattice vector $2\vec{\tau}$ if $\vec{Q}$ is equivalent to
$C\vec{Q}$ up to a reciprocal lattice vector, $\tilde{f}_{A}(\vec{r})+\tilde{f}_{B}(\vec{r})$
is odd under inversion when $e^{-i2\vec{Q}\cdot\vec{\tau}}=-1$. Consequently,
$f_{A}(\vec{r})+f_{B}(\vec{r})$ and $\tilde{f}_{A}(\vec{r})+\tilde{f}_{B}(\vec{r})$
are linearly independent and related by symmetry but have opposite
parity, and thus they should be used together to form a basis for
an irreducible representation. The resultant irreducible representation
cannot be assigned a definite parity.

\subsection{Sublattice-preserving inversion}

If a WP possesses an inversion symmetry $\{{\cal I}|\vec{t}\}$, which
may be accompanied by a lattice translation $\vec{t}$, then there
should be another symmetry $\{C|\vec{\tau}\}$ mapping sublattice
$A$ to $B$. With these symmetries, we can show $\{C|\vec{\tau}\}\{{\cal I}|\vec{t}\}=\{e|2\vec{\tau}-(1-C)\vec{t}\}\{{\cal I}|\vec{t}\}\{C|\vec{\tau}\}$
along with $\vec{t}=2\vec{r}_{A}$ and $\vec{r}_{B}=\vec{\tau}+C\vec{r}_{A}$.
Taking one of the sublattices as the origin, i.e. $\vec{r}_{A}=0$,
we can set $\vec{t}=0$ which leads to a simpler result: 
\begin{equation}
\{C|\vec{\tau}\}\{{\cal I}|0\}=\{e|2\vec{\tau}\}\{{\cal I}|0\}\{C|\vec{\tau}\},\label{eq:Sublattice_Preserving_Case}
\end{equation}
with $\vec{r}_{B}=\vec{\tau}$. Note that $2\vec{\tau}$ should be
a lattice vector, and the requirement $2(C\vec{Q})\cdot\vec{\tau}\in(2\mathbb{Z}+1)\pi$
for a half reciprocal lattice vector $\vec{Q}$ implies that $\vec{\tau}$
has to be a half-integer translation vector.

Having inversion as a site symmetry, we can construct even-parity
and odd-parity functions $f_{A}^{(e)}(\vec{r})$ and $f_{A}^{(o)}(\vec{r})$,
which can be thought to be finite on the lattice of the sublattice
$A$, and oscillating with a propagation vector $\vec{Q}$. Then,
$\{C|\vec{\tau}\}f_{A}^{(p)}(\vec{r})$ ($p=e,o$) has a finite amplitudes
on the lattice of sublattice $B$. Furthermore, Eq. \eqref{eq:Sublattice_Preserving_Case}
dictates that the parity of $\{C|\vec{\tau}\}f_{A}^{(p)}(\vec{r})$ is opposite to $f_{A}^{(p)}(\vec{r})$ if $e^{2i\vec{Q}\cdot\vec{\tau}}=-1$,
as it can be known as follows:
\begin{align}
\{{\cal I}|0\}\{C|\vec{\tau}\}f_{A}^{(p)}(\vec{r})= & \{e|-2\vec{\tau}\}\{C|\vec{\tau}\}\{{\cal I}|0\}f_{A}^{(p)}(\vec{r})\notag \\
= & \lambda_{p}e^{2i\vec{Q}\cdot\vec{\tau}}\{C|\vec{\tau}\}f_{A}^{(p)}(\vec{r}),
\end{align}
where $\lambda_{p}=\pm1$ ($p=e$, $o$) is the parity of the function
$f_{A}^{(p)}$. Thus, $f_{A}^{(p)}(\vec{r})$ and $\{C|\vec{\tau}\}f_{A}^{(p)}(\vec{r})$
are linearly independent, and should be taken into account together
to construct a set of basis functions for an irreducible representation
of the space group. The resultant irreducible representation made
from $f_{A}^{(p)}(\vec{r})$ and $\{C|\vec{\tau}\}f_{A}^{(p)}(\vec{r})$
involves functions of opposite parities.

\section{Symmetry properties of $\vec{S}_{A}$ and $\vec{S}_{B}$, and the
forbidden terms in Landau free energy}

\label{App:OddPower_SA_SB}In this section, we expose the symmetry
properties the antiferromagnetic order parameters $\vec{S}_{A}$ and
$\vec{S}_{B}$. Though it is possible to proceed with the basis functions
$f_{A}(\vec{r})$ and $f_{B}(\vec{r})$ introduced in Sec. \ref{App:MixedRep},
we take a group theoretical approach to put it on a more general ground. 

\subsection{Symmetry properties of $\vec{S}_{A}$ and $\vec{S}_{B}$}

The inversion symmetry has a different role depending on the type of
the Wyckoff position, and we first begin with the case of type-1 Wyckoff
position, where the inversion exchanges the two sublattices. 

The symmetry properties of the antiferromagnetic order parameters $\vec{S}_{A}$
and $\vec{S}_{B}$ can be examined by considering a vector field 
\begin{equation}
\vec{S}(\vec{r})=\vec{S}_{A}f_{A}(\vec{r})+\vec{S}_{B}f_{B}(\vec{r}),\label{eq:vectorfieldS}
\end{equation}
where $f_{B}(\vec{r})$ is defined as $\{{\cal I}|0\}f_{A}(\vec{r})$
with $f_{A}(\vec{r}+\vec{R})=e^{i\vec{Q}\cdot\vec{R}}f_{A}(\vec{r})$.
The functions in Eqs.~(\ref{eq:fA}--\ref{eq:fB}) can serve as
$f_{A}(\vec{r})$ and $f_{B}(\vec{r})$ in Eq.~\eqref{eq:vectorfieldS},
but the following argument does not rely on the
specific form of these functions. As $\{{\cal I}|0\}^{2}$ is the identity
operator $\{e|0\}$, we have 
\begin{equation}
\{{\cal I}|0\}f_{B}(\vec{r})=f_{A}(\vec{r}).
\end{equation}

Furthermore, for a symmetry leaving the site $\vec{r}_{A}$ unmoved,
we can assume $\{C|\vec{\tau}\}f_{A}(\vec{r})=\lambda_{C}f_{A}(\vec{r})$
with $\lambda_{C}=\pm1$, regardless of
whether $f_{A}(\vec{r})$ has its maximum on the sublattice $A$ or
not. Without loss of generality, we take $\lambda_{C}=1$; the discussion below can be straightforwardly applied to the case $\lambda_C=-1$. Given the group theoretical identity $\{C|\vec{\tau}\}\{{\cal I}|0\}=\{e|2\vec{\tau}\}\{{\cal I}|0\}\{C|\vec{\tau}\}$,
we can show
\begin{align}
\{C|\vec{\tau}\}f_{B} & =\{C|\vec{\tau}\}\{{\cal I}|0\}f_{A}\nonumber \\
 & =\{e|2\vec{\tau}\}\{{\cal I}|0\}\{C|\vec{\tau}\}f_{A}\nonumber \\
 & =\{e|2\vec{\tau}\}f_{B}\nonumber \\
 & = e^{-2i\vec{Q}\cdot\vec{\tau}}f_{B}=-f_{B}
\end{align}
These symmetry transformation rules of the basis functions $f_{A}$
and $f_{B}$ induces the representation matrices of $\{{\cal I}|0\}$
and $\{C|\vec{\tau}\}$ in the basis of $\vec{S}_{A}$ and $\vec{S}_{B}$,
which can be represented as follows:
\begin{widetext}
\begin{equation}
\{{\cal I}|0\}\left(\begin{matrix}\vec{S}_{A}\\
\vec{S}_{B}
\end{matrix}\right)=\left(\begin{matrix} & 1\\
1
\end{matrix}\right)\left(\begin{matrix}\vec{S}_{A}\\
\vec{S}_{B}
\end{matrix}\right),\quad\{C|\vec{\tau}\}\left(\begin{matrix}\vec{S}_{A}\\
\vec{S}_{B}
\end{matrix}\right)=\left(\begin{matrix}1\\
 & -1
\end{matrix}\right)\left(\begin{matrix}\vec{S}_{A}\\
\vec{S}_{B}
\end{matrix}\right).\label{eq:symrule_type1}
\end{equation}
\end{widetext}

When it comes to the type-2 Wyckoff position, we define $f_{B}(\vec{r})\equiv\{C|\vec{\tau}\}f_{A}(\vec{r})$,
where $f_{A}(\vec{r})$ is assumed to obey $\{{\cal I}|0\}f_{A}(\vec{r})=\lambda_{{\cal I}}f_{A}(\vec{r})$
with $\lambda_{{\cal I}}=\pm1$. Without loss of generality, we take
$\lambda_{{\cal I}}=+1$ in the following presentation. The group
theoretical identity $\{C|\vec{\tau}\}\{{\cal I}|0\}=\{e|2\vec{\tau}\}\{{\cal I}|0\}\{C|\vec{\tau}\}$
results in 
\begin{align}
\{{\cal I}|0\}f_{B} & =\{{\cal I}|0\}\{C|\vec{\tau}\}f_{A}\nonumber \\
 & =\{e|-2\vec{\tau}\}\{C|\vec{\tau}\}\{{\cal I}|0\}f_{A}\nonumber \\
 & =e^{2i\vec{Q}\cdot\vec{\tau}}f_{B}
 =-f_{B}
 .
\end{align}
 Furthermore, as $\{C|\vec{\tau}\}$ exchanges two sublattices, $\{C|\vec{\tau}\}^{2}$
should preserve a sublattice. In other words, it maps a site $\vec{r}_{A}$
to another site equivalent to $\vec{r}_{A}$ up to a lattice vector.
Therefore, $\{C|\vec{\tau}\}f_{B}(\vec{r})\equiv\{C|\vec{\tau}\}^{2}f_{A}(\vec{r})=\{C^{2}|C\vec{\tau}+\vec{\tau}\}f_{A}(\vec{r})=e^{-iC^{-2}\vec{Q}\cdot(C\vec{\tau}+\vec{\tau})}f_{A}(\vec{r})$.
Note that it is generally possible to identify $\vec{r}_{A}$ with
the origin, and the equivalence of $\{C^{2}|C\vec{\tau}+\vec{\tau}\}\vec{r}_{A}=C^{2}\vec{r}_{A}+C\vec{\tau}+\vec{\tau}$
and $\vec{r}_{A}=(0,0,0)$ means that $C\vec{\tau}+\vec{\tau}$ is
a translation. As a result, we have $e^{-iC^{-2}\vec{Q}\cdot(C\vec{\tau}+\vec{\tau})}=e^{-i\vec{Q}\cdot(C\vec{\tau}+\vec{\tau})}=\pm1$.
Whether $e^{-i\vec{Q}\cdot(C\vec{\tau}+\vec{\tau})}=+1$ or $e^{-i\vec{Q}\cdot(C\vec{\tau}+\vec{\tau})}=-1$
varies from case to case. 

Given the trnasformation rule of $f_{A}$ and $f_{B}$ under $\{{\cal I}|0\}$
and $\{C|\vec{\tau}\}$, we can obtain the symmetry transformation
rule of $\vec{S}_{A}$ and $\vec{S}_{B}$ under those two symmetries,
which are represented as

\begin{widetext}

\begin{equation}
\{{\cal I}|0\}\left(\begin{matrix}\vec{S}_{A}\\
\vec{S}_{B}
\end{matrix}\right)=\left(\begin{matrix}1\\
 & -1
\end{matrix}\right)\left(\begin{matrix}\vec{S}_{A}\\
\vec{S}_{B}
\end{matrix}\right),\quad\{C|\vec{\tau}\}\left(\begin{matrix}\vec{S}_{A}\\
\vec{S}_{B}
\end{matrix}\right)=\left(\begin{matrix} & e^{-i\vec{Q}\cdot(C\vec{\tau}+\vec{\tau})}\\
1
\end{matrix}\right)\left(\begin{matrix}\vec{S}_{A}\\
\vec{S}_{B}
\end{matrix}\right).\label{eq:symrule_type2}
\end{equation}
\end{widetext}

The symmetry transformation rules of $\vec{S}_{A}$ and $\vec{S}_{B}$
given in Eq.~\eqref{eq:symrule_type1} and Eq.~\eqref{eq:symrule_type2}
are enough to rule out the terms $\vec{S}_{A}\cdot\vec{S}_{B}$, $|\vec{S}_{A}|^{2}-|\vec{S}_{B}|^{2}$,
$(|\vec{S}_{A}|^{2}+|\vec{S}_{B}|^{2})\vec{S}_{A}\cdot\vec{S}_{B}$,
and $(|\vec{S}_{A}|^{2}-|\vec{S}_{B}|^{2})\vec{S}_{A}\cdot\vec{S}_{B}$
in the Landau free energy in the main text in both type-1 Wyckoff
position case and the type-2 Wyckoff position.

\section{Symmetry propeties of $\vec{t}_{\vec{k}}$}

\label{App:sym_t}

\subsection{In the conventional gauge}

As discussed in the main text, the hopping vector $\vec{t}_{\vec{k}}$
plays the important role in judging the possiblity of OP-AFM phase
with spin-split bands. Especially, having rather small $|\vec{t}_{\vec{k}}\times\vec{t}_{\vec{k}+\vec{Q}}|$
is favorable for the emergence of the OP-AFM phase. Depending on the
dimensionality of the bands, such condition can be ensured by symmetry. 

In the gauge that the Bloch basis function is periodic in the Brillouin
zone, the transformation for the fermionic annihiliaton operator by
a symmetry $\hat{g}=\{R|\vec{\tau}_{R}\}$ is given by $\hat{g}^{-1}\hat{C}_{\vec{k}}\hat{g}=\Gamma_{\vec{k}}(g)\hat{C}_{R^{-1}\vec{k}}$
with 
\begin{equation}
\Gamma_{\vec{k}}(g)=e^{-i\vec{k}\cdot\vec{\tau}}V_{\vec{k}}\Gamma(g)V_{R^{-1}\vec{k}}^{\dagger}.
\end{equation}
where $[V_{\vec{k}}]_{ij}=\delta_{ij}e^{i\vec{k}\cdot\vec{r}_{i}}$
and $\Gamma(g)$ is a $\vec{k}$-independent matrix. $\vec{r}_{i}$
denotes the position of the atoms within a unit cell. We foucs on
the cases with two sublattices at $\vec{r}_{A}$ and $\vec{r}_{B}$
which provide electrons with $s$-wave-like orbital character. In this circumstatnce,
we have $\Gamma(g)\propto\tau_{0}$ or $\Gamma(g)\propto\tau_{x}$
depending on whether the symmetry $g$ belongs to a site symmetry
of a sublattice or not. 

The resultant symmetry properties of $\vec{t}_{\vec{k}}=(t_{\vec{k},x},t_{\vec{k},y})$
can be expressed as \begin{widetext}
\begin{align}
t_{\vec{k},x}-it_{\vec{k},y}= & \begin{cases}
e^{i(R^{-1}-1)\vec{k}\cdot(r_{A}-r_{B})}(t_{R\vec{k},x}-it_{R\vec{k},y}) & g\in\text{site symmetry group of \ensuremath{A}}\\
e^{i(R^{-1}+1)\vec{k}\cdot(r_{A}-r_{B})}(t_{R\vec{k},x}+it_{R\vec{k},y}) & g\notin\text{site symmetry group of \ensuremath{A}}
\end{cases}.
\end{align}
If a site-symmetry $\{S|\vec{\tau}_{S}\}$ of the sublattice $B$
leaving $\vec{k}$ invariant up to a reciprocal lattice vector is
present and satisfies $e^{-2i\vec{k}\cdot\tau_{S}}=-1$, the first
line of the above equation implies  is the vanising of $t_{\vec{k},x}$
and $t_{\vec{k},y}$ on such $\vec{k}$. In addition to the spatial
symmetries, the time-reversal symmetry ${\cal T}$ provides additional
rule
\begin{align}
t_{\vec{k},x}-it_{\vec{k},y} & =t_{-\vec{k},x}+it_{-\vec{k},y}.
\end{align}

\subsubsection{Type-1 class}

In the type-1 class, $\{{\cal I}|0\}$ is a sublattice-transposing
symmetry, and we take $\{C|\vec{\tau}\}$ as the sublattice-preserving
one with $e^{2i\vec{Q}\cdot\vec{\tau}}=-1$. Furthermore, $C\vec{r}_{A}-\vec{r}_{A}=\vec{\tau}$
and $C\vec{r}_{B}-\vec{r}_{B}=-\vec{\tau}$ as $\vec{r}_{B}=-\vec{r}_{A}$.
This leads to $e^{i(C^{-1}-1)\vec{k}\cdot(r_{A}-r_{B})}=e^{2i\vec{k}\cdot\vec{\tau}}$
and we obtain

\begin{align}
t_{\vec{k},x}-it_{\vec{k},y}= & \begin{cases}
e^{2i\vec{k}\cdot\vec{\tau}}(t_{C\vec{k},x}-it_{C\vec{k},y}) & g=\{C|\vec{\tau}\},\\
t_{-\vec{k},x}+it_{-\vec{k},y} & g=\{{\cal I}|0\}\;{\rm and}\;{\cal T},
\end{cases}
\end{align}
where the second line can also be obtained through the time-reversal
symmetry. Using these relations, we obtain the following:
\[
(t_{\vec{k},x}-it_{\vec{k},y})(t_{\vec{k}+\vec{Q},x}+it_{\vec{k}+\vec{Q},y})=\begin{cases}
e^{2i\vec{Q}\cdot\vec{\tau}}(t_{C\vec{k},x}-it_{C\vec{k},y})(t_{C\vec{k}+\vec{Q},x}+it_{C\vec{k}+\vec{Q},y}) & \{C|\vec{\tau}\},\\
(t_{-\vec{k},x}+it_{-\vec{k},y})(t_{-\vec{k}+\vec{Q},x}-it_{-\vec{k}+\vec{Q},y}) & \{{\cal I}|0\}\;{\rm and}\;{\cal T},
\end{cases}
\]
with $e^{2i\vec{Q}\cdot\vec{\tau}}=-1$.

The symmetry properties of $\vec{t}_{\vec{k}}\cdot\vec{t}_{\vec{k}+\vec{Q}}+i(\vec{t}_{\vec{k}}\times\vec{t}_{\vec{k}+\vec{Q}})=(t_{\vec{k},x}-it_{\vec{k},y})(t_{\vec{k}+\vec{Q},x}+it_{\vec{k}+\vec{Q},y})$
are given by 
\begin{align}
(t_{\vec{k},x}-it_{\vec{k},y})(t_{\vec{k}+\vec{Q},x}+it_{\vec{k}+\vec{Q},y}) & =-(t_{C\vec{k},x}-it_{C\vec{k},y})(t_{C\vec{k}+\vec{Q},x}+it_{C\vec{k}+\vec{Q},y}),\label{eq:SublatticePreserving}
\end{align}
due to a symmetry $\{C|\vec{\tau}\}$, and 
\begin{align}
(t_{\vec{k},x}-it_{\vec{k},y})(t_{\vec{k}+\vec{Q},x}+it_{\vec{k}+\vec{Q},y}) & =(t_{-\vec{k},x}+it_{-\vec{k},y})(t_{-\vec{k}+\vec{Q},x}-it_{-\vec{k}+\vec{Q},y}),\label{eq:SublatticeTransposing}
\end{align}
due to symmetries $\{{\cal I}|0\}$ and ${\cal T}$. 

Besides the effect of the two primary symmetries $\{{\cal I}|0\}$
and $\{C|\vec{\tau}\}$, additional symmetries can produce important
consequences, too. Supposing that there is a sublattice-preserving
symmetry $\{R|\vec{\tau}_{R}\}$ leaving $\vec{Q}$ invariant and
satisfying $e^{2i\vec{Q}\cdot\vec{\tau}_{R}}=+1$, we obtain 
\[
(t_{\vec{k},x}-it_{\vec{k},y})(t_{\vec{k}+\vec{Q},x}+it_{\vec{k}+\vec{Q},y})=+(t_{R\vec{k},x}-it_{R\vec{k},y})(t_{R\vec{k}+\vec{Q},x}+it_{R\vec{k}+\vec{Q},y}).
\]
Combined with the inversion or the time-reversal, this makes $\vec{t}_{\vec{k}}\times\vec{t}_{\vec{k}+\vec{Q}}$
vanish on $\vec{k}$ satisfying $\vec{k}\equiv-R\vec{k}$ up to a
reciprocal lattice vector. For example, let us consider the space
group~129 where all Wykcoff positions of multiplicty 2 are of
type-1. In this system, the two-fold rotaion $\{C_{2z}|\frac{1}{2}\frac{1}{2}0\}$
is a sublattice-preserving symmetry of the type-1 WPs and leaves $\vec{Q}=(\pi,\pi,0)$
and $\vec{Q}=(\pi,\pi,\pi)$ invariant upto a reciprocal lattice vector
as well as satisfying $e^{2i\vec{Q}\cdot\vec{\tau}_{R}}=+1$. When
combined with the time-reversal symmetry, Eq.~\eqref{eq:SublatticePreserving}
leads to $\vec{t}_{\vec{k}}\times\vec{t}_{\vec{k}+\vec{Q}}=0$ on
the $k_{z}=0$ plane. Therefore, if the strong nesting between the
bands $\xi_{\vec{k},\pm}$ and $\xi_{\vec{k}+\vec{Q},\pm}$ occurs
in this plane, $\beta_{2}$ will be positive, and OP-AFM phase is
expected to emerge with spin-split bands at general $k_{z}$. 

\subsubsection{Type-2 class}

In the type-2 class, $\{{\cal I}|0\}$ is a sublattice-preserving
symmetry, and we take $\{C|\vec{\tau}\}$ as the sublattice-transposing
one. This let us to set $\vec{r}_{A}=0$ and $\vec{r}_{B}=\vec{\tau}$
and we get

\begin{align}
t_{\vec{k},x}-it_{\vec{k},y}= & \begin{cases}
e^{2i\vec{k}\cdot\vec{\tau}}(t_{-\vec{k},x}-it_{-\vec{k},y}) & \{{\cal I}|0\},\\
e^{i(C^{-1}+I)\vec{k}\cdot(r_{A}-r_{B})}(t_{C\vec{k},x}+it_{C\vec{k},y}) & \{C|\vec{\tau}\},\\
t_{-\vec{k},x}+it_{-\vec{k},y} & {\cal T},
\end{cases}
\end{align}
which results in 
\begin{equation}
(t_{\vec{k},x}-it_{\vec{k},y})(t_{\vec{k}+\vec{Q},x}+it_{\vec{k}+\vec{Q},y})=\begin{cases}
-(t_{-\vec{k},x}-it_{-\vec{k},y})(t_{-\vec{k}+\vec{Q},x}+it_{-\vec{k}+\vec{Q},y}) & \{{\cal I}|0\},\\
e^{i\rho}(t_{C\vec{k},x}+it_{C\vec{k},y})(t_{C\vec{k}+\vec{Q},x}-it_{C\vec{k}+\vec{Q},y}) & \{C|\vec{\tau}\},\\
(t_{-\vec{k},x}+it_{-\vec{k},y})(t_{-\vec{k}+\vec{Q},x}-it_{-\vec{k}+\vec{Q},y}) & {\cal T},
\end{cases}
\end{equation}
with $\rho=\vec{Q}\cdot(C\vec{\tau}+\vec{\tau})$. Note that the space-time
inversion symmetry dictates that $\vec{t}_{\vec{k}}\cdot\vec{t}_{\vec{k}+\vec{Q}}$
is identically \emph{zero} \emph{for} \emph{all} $\vec{k}$ in type-2
systems.

Regarding $\varepsilon_{\vec{k},z}$ which exists only when the inversion
is a site symmetry of a sublattice, its transformation rule under
a sublattice-transposing symmetry $g$ is given by
\begin{equation}
\varepsilon_{R\vec{k},z}=-\varepsilon_{\vec{k},z}.
\end{equation}

\begin{table*}
\begin{tabular}{|c|c|c|c|c|c|c|}
\hline 
 & \multicolumn{3}{c|}{Type-1} & \multicolumn{3}{c|}{Type-2}\tabularnewline
\hline 
 & $\{C|\vec{\tau}\}$ & $\{{\cal I}|0\}$ $({\cal T})$ & $\{R|\vec{\tau}_{R}\}$ & $\{{\cal I}|0\}$ $({\cal T})$ & $\{C|\vec{\tau}\}$ & $\{R|\vec{\tau}_{R}\}$\tabularnewline
\hline 
\hline 
$\vec{t}_{\vec{k}}\cdot\vec{t}_{\vec{k}+\vec{Q}}$ & $-\vec{t}_{C\vec{k}}\cdot\vec{t}_{C\vec{k}+\vec{Q}}$ & $\vec{t}_{-\vec{k}}\cdot\vec{t}_{-\vec{k}+\vec{Q}}$ & $\vec{t}_{R\vec{k}}\cdot\vec{t}_{R\vec{k}+\vec{Q}}$ & $-(+)\vec{t}_{-\vec{k}}\cdot\vec{t}_{-\vec{k}+\vec{Q}}$ & $-e^{i\rho}\vec{t}_{C\vec{k}}\cdot\vec{t}_{C\vec{k}+\vec{Q}}$ & $\vec{t}_{R\vec{k}}\cdot\vec{t}_{R\vec{k}+\vec{Q}}$\tabularnewline
\hline 
$\vec{t}_{\vec{k}}\times\vec{t}_{\vec{k}+\vec{Q}}$ & $-\vec{t}_{C\vec{k}}\times\vec{t}_{C\vec{k}+\vec{Q}}$ & $-\vec{t}_{-\vec{k}}\times\vec{t}_{-\vec{k}+\vec{Q}}$ & $\vec{t}_{R\vec{k}}\times\vec{t}_{R\vec{k}+\vec{Q}}$ & $-\vec{t}_{-\vec{k}}\times\vec{t}_{-\vec{k}+\vec{Q}}$ & $e^{i\rho}\vec{t}_{C\vec{k}}\times\vec{t}_{C\vec{k}+\vec{Q}}$ & $\vec{t}_{R\vec{k}}\times\vec{t}_{R\vec{k}+\vec{Q}}$\tabularnewline
\hline 
$\varepsilon_{\vec{k},z}$ & -- & -- & -- & $\varepsilon_{-\vec{k},z}$ & $-\varepsilon_{C\vec{k},z}$ & $\varepsilon_{R\vec{k},z}$\tabularnewline
\hline 
\end{tabular}\caption{\label{Apptab:t_conventional}The symmetry transformation rules under
three types of symmetries for $\vec{Q}$. For the case where the parities
under the inversion and the time-reversal symmetries are different,
the parity under the time-reversal is denoted in parenthesis. $\{R|\vec{\tau}_{R}\}$
is an additional symmetry leaving $\vec{Q}$ invariant and satisfying
$e^{2i\vec{Q}\cdot\vec{\tau}_{R}}=+1$. $\rho=\vec{Q}\cdot(C\vec{\tau}-\vec{\tau})$.
$\varepsilon_{\vec{k},z}$ is absent for the type-1 case.}
\end{table*}

\subsection{In the symmetric gauge}

Here, for convenience in dealing with symmetry operation, we introduce
a gauge where the Bloch basis of electronic wave functions are defined
as 
\[
\hat{\tilde{C}}_{\vec{k},\tau,\sigma}=\frac{1}{N}\sum_{\vec{R}}e^{i\vec{k}\cdot(\vec{R}+\vec{\tilde{r}}_{\tau})}\hat{C}_{\vec{R},\tau,\sigma}
\]
where $\vec{\tilde{r}}_{\tau}$ ($\tau=A,B$) correspond the flattened
position of the sublattices with unfixed coordinates are flattened
to zero. For example, the Wykcoff position $2c$ in the space group
No.~129 includes two sublattices at $\vec{r}_{A}=(1/4,1/4,z)$ and
$\vec{r}_{B}=(3/4,3/4,-z)$, and $\vec{\tilde{r}}_{A}=(1/4,1/4,0)$
and $\vec{\tilde{r}}_{A}=(3/4,3/4,0)$ are the corresponding flattened
coordinates. As one can see, this does not affect the cases where
all the three coordinates of $\vec{r}_{\tau}$ are fixed by symmetry,
which happens when an inversion symmetry is a site symmetry of a sublattice.
This gauge is particulary useful when dealing with the symmetry operation
because the representation matrix of symmetry appears not to depend
on the momentum unlike the conventional gauge featuring the periodicity
of the Hamiltonian in the momentum space. 

In this gauge with flattened coordinates of sublattices, the transformation
for the fermionic annihiliaton operator by a symmetry $\hat{g}=\{R|\vec{\tau}\}$
is given simply by $\hat{g}^{-1}\hat{\tilde{C}}_{\vec{k}}\hat{g}=e^{-i\vec{k}\cdot\vec{\tau}}\Gamma(g)\hat{\tilde{C}}_{R^{-1}\vec{k}}$
with $\Gamma(g)$ not depending on the momentum. For a sublattice-preserving
and a sublattice-transposing symmetries, $\Gamma(g)\propto\tau_{0}$
and $\Gamma(g)\propto\tau_{x}$, respectively. As a result, we obtain

\begin{align}
(\tilde{t}_{R^{-1}\vec{k},x},\tilde{t}_{R^{-1}\vec{k},y},\varepsilon_{R^{-1}\vec{k},z})= & \begin{cases}
(\tilde{t}_{\vec{k},x},\tilde{t}_{\vec{k},y},\varepsilon_{\vec{k},z}) & \{R|\vec{\tau}\}\in\text{site symmetry group of \ensuremath{A}}\\
(\tilde{t}_{\vec{k},x},-\tilde{t}_{\vec{k},y},-\varepsilon_{\vec{k},z}) & \{R|\vec{\tau}\}\notin\text{site symmetry group of \ensuremath{A}}
\end{cases}.
\end{align}
 $\tilde{t}_{\vec{k},x}-i\tilde{t}_{\vec{k},y}$ is related with $t_{\vec{k},x}-it_{\vec{k},y}$
through $\tilde{t}_{\vec{k},x}-i\tilde{t}_{\vec{k},y}=(t_{\vec{k},x}-it_{\vec{k},x})e^{i\vec{k}(\vec{r}_{A}-\vec{r}_{B})}$
from which we derive
\begin{align*}
\vec{\tilde{t}}_{\vec{k}}\cdot\vec{\tilde{t}}_{\vec{k}+\vec{Q}}= & {\rm Re}[(t_{\vec{k},x}-it_{\vec{k},y})(t_{\vec{k}+\vec{Q},x}+it_{\vec{k}+\vec{Q},x})e^{i\rho}]\\
= & \cos\rho\cdot(\vec{t}_{\vec{k}}\cdot\vec{t}_{\vec{k}+\vec{Q}})-\sin\rho(\vec{t}_{\vec{k}}\times\vec{t}_{\vec{k}+\vec{Q}})_{z},\\
(\vec{\tilde{t}}_{\vec{k}}\times\vec{\tilde{t}}_{\vec{k}+\vec{Q}})_{z}= & {\rm Im}[(t_{\vec{k},x}-it_{\vec{k},y})(t_{\vec{k}+\vec{Q},x}+it_{\vec{k}+\vec{Q},x})e^{i\rho}]\\
= & \sin\rho\cdot(\vec{t}_{\vec{k}}\cdot\vec{t}_{\vec{k}+\vec{Q}})+\cos\rho(\vec{t}_{\vec{k}}\times\vec{t}_{\vec{k}+\vec{Q}})_{z},
\end{align*}
with $\rho=\vec{Q}\cdot(\vec{r}_{B}-\vec{r}_{A})$. As $R\vec{Q}\cdot(\vec{r}_{B}-\vec{r}_{A})\neq\vec{Q}\cdot(\vec{r}_{B}-\vec{r}_{A})$
in general, this should be taken into account when we derive the transformation
rule for $(\tilde{t}_{\vec{k}+\vec{Q},x},\tilde{t}_{\vec{k}+\vec{Q},y})$.
$\tilde{t}_{\vec{k}+R^{-1}\vec{Q},x}-i\tilde{t}_{\vec{k}+R^{-1}\vec{Q},y}$
and $\tilde{t}_{\vec{k}+\vec{Q},x}-i\tilde{t}_{\vec{k}+\vec{Q},y}$
are connected through 
\[
\tilde{t}_{\vec{k}+R^{-1}\vec{Q},x}-i\tilde{t}_{\vec{k}+R^{-1}\vec{Q},y}=(\tilde{t}_{\vec{k}+\vec{Q},x}-i\tilde{t}_{\vec{k}+\vec{Q},y})e^{i(R^{-1}\vec{Q}-\vec{Q})\cdot(\vec{\tilde{r}}_{B}-\vec{\tilde{r}}_{A})}.
\]
We note that $e^{i(R^{-1}\vec{Q}-\vec{Q})\cdot(\vec{\tilde{r}}_{B}-\vec{\tilde{r}}_{A})}=-1$
for a sublattice-preserving symmetry $g=\{R|\vec{\tau}\}$. This can
be proved as following:
\begin{itemize}
\item In the type-1 case, we have $C\vec{r}_{A}+\vec{\tau}=\vec{r}_{A}$
and $C\vec{r}_{B}-\vec{\tau}=\vec{r}_{B}$. This leads to $(C-1)(\vec{r}_{B}-\vec{r}_{A})=-2\vec{\tau}$,
and thus we obtain $e^{i(C^{-1}\vec{Q}-\vec{Q})\cdot(\vec{r}_{B}-\vec{r}_{A})}=-1$. 
\item In the type-2 case, we can set $\vec{r}_{A}=0$ and $\vec{r}_{B}=\vec{\tau}$
due to a sublattice-transposing symmetry $\{C|\vec{\tau}\}$. As a
result, we get $e^{i({\cal I}^{-1}\vec{Q}-\vec{Q})\cdot(\vec{r}_{B}-\vec{r}_{A})}=e^{-2i\vec{Q}\cdot\vec{\tau}}=-1.$
\end{itemize}
Consequently, $(\tilde{t}_{\vec{k}+\vec{Q},x},\tilde{t}_{\vec{k}+\vec{Q},y},\varepsilon_{\vec{k}+\vec{Q},z})$
transform under a sublattice-preserving symmetry $g=\{R|\vec{\tau}\}$
as 
\begin{equation}
(\tilde{t}_{R^{-1}\vec{k}+\vec{Q},x},\tilde{t}_{R^{-1}\vec{k}+\vec{Q},y},\varepsilon_{R^{-1}\vec{k}+\vec{Q},z})=(-\tilde{t}_{\vec{k}+\vec{Q},x},-\tilde{t}_{\vec{k}+\vec{Q},y},\varepsilon_{\vec{k}+\vec{Q},z}).
\end{equation}
The symmetry properties of $(\tilde{t}_{\vec{k},x},\tilde{t}_{\vec{k},y},\varepsilon_{\vec{k},z})$
and $(\tilde{t}_{\vec{k}+\vec{Q},x},\tilde{t}_{\vec{k}+\vec{Q},y},\varepsilon_{\vec{k}+\vec{Q},z})$
dictates that both $\vec{\tilde{t}}_{\vec{k}}\cdot\vec{\tilde{t}}_{\vec{k}+\vec{Q}}$
and $\vec{\tilde{t}}_{\vec{k}}\times\vec{\tilde{t}}_{\vec{k}+\vec{Q}}$
are odd under the sublattice-preserving symmetry, while $\varepsilon_{\vec{k},z}\varepsilon_{\vec{k}+\vec{Q},z}$
is odd under the sublattice-transposing symmetry.

Consequently, $(\tilde{t}_{\vec{k}+\vec{Q},x},\tilde{t}_{\vec{k}+\vec{Q},y},\varepsilon_{\vec{k}+\vec{Q},z})$
transform under a sublattice-preserving symmetry $g=\{R|\vec{\tau}\}$
as 
\begin{equation}
(\tilde{t}_{R^{-1}\vec{k}+\vec{Q},x},\tilde{t}_{R^{-1}\vec{k}+\vec{Q},y},\varepsilon_{R^{-1}\vec{k}+\vec{Q},z})=(-\tilde{t}_{\vec{k}+\vec{Q},x},-\tilde{t}_{\vec{k}+\vec{Q},y},\varepsilon_{\vec{k}+\vec{Q},z}).
\end{equation}
The symmetry properties of $(\tilde{t}_{\vec{k},x},\tilde{t}_{\vec{k},y},\varepsilon_{\vec{k},z})$
and $(\tilde{t}_{\vec{k}+\vec{Q},x},\tilde{t}_{\vec{k}+\vec{Q},y},\varepsilon_{\vec{k}+\vec{Q},z})$
dictates that both $\vec{\tilde{t}}_{\vec{k}}\cdot\vec{\tilde{t}}_{\vec{k}+\vec{Q}}$
and $\vec{\tilde{t}}_{\vec{k}}\times\vec{\tilde{t}}_{\vec{k}+\vec{Q}}$
are odd under the sublattice-preserving symmetry, while $\varepsilon_{\vec{k},z}\varepsilon_{\vec{k}+\vec{Q},z}$
is odd under the sublattice-transposing symmetry.

The transformation rule under the time-reversal symmetry is given
by

\begin{align}
(\tilde{t}_{-\vec{k},x},\tilde{t}_{-\vec{k},y},\varepsilon_{-\vec{k},z})= & (\tilde{t}_{\vec{k},x},-\tilde{t}_{\vec{k},y},\varepsilon_{\vec{k},z}),\\
(\tilde{t}_{-\vec{k}+\vec{Q},x},\tilde{t}_{-\vec{k}+\vec{Q},y},\varepsilon_{-\vec{k}+\vec{Q},z})= & e^{-2i\rho}(\tilde{t}_{\vec{k}+\vec{Q},x},-\tilde{t}_{\vec{k}+\vec{Q},y},\varepsilon_{\vec{k}+\vec{Q},z}),
\end{align}
with $\rho=\vec{Q}\cdot(\vec{\tilde{r}}_{B}-\vec{\tilde{r}}_{A})$.
For the type-2 case, $e^{-2i\rho}=-1$, which means that $\vec{\tilde{t}}_{\vec{k}}\cdot\vec{\tilde{t}}_{\vec{k}+\vec{Q}}$
is odd under the time-reversal symmetry while $\vec{\tilde{t}}_{\vec{k}}\times\vec{\tilde{t}}_{\vec{k}+\vec{Q}}$
and $\tilde{t}_{\vec{k},z}\tilde{t}_{\vec{k}+\vec{Q},z}$ are even.

Table~\ref{Apptab:t_hybrid} summaries the result. Note that the inversion
and the time reversal symmetries make $\vec{\tilde{t}}_{\vec{k}}\times\vec{\tilde{t}}_{\vec{k}+\vec{Q}}=0$
in the type-2 case. 

\begin{table*}
\begin{tabular}{|c|c|c|c|c|c|c|}
\hline 
 & \multicolumn{3}{c|}{Type-1} & \multicolumn{3}{c|}{Type-2}\tabularnewline
\hline 
 & $\{C|\vec{\tau}\}$ & $\{{\cal I}|0\}$ & ${\cal T}$ & $\{{\cal I}|0\}$ & $\{C|\vec{\tau}\}$ & ${\cal T}$\tabularnewline
\hline 
\hline 
$\vec{\tilde{t}}_{\vec{k}}\cdot\vec{\tilde{t}}_{\vec{k}+\vec{Q}}$ & $-\vec{\tilde{t}}_{C\vec{k}}\cdot\vec{\tilde{t}}_{C\vec{k}+\vec{Q}}$ & $e^{-2i\rho}\vec{\tilde{t}}_{-\vec{k}}\cdot\vec{\tilde{t}}_{-\vec{k}+\vec{Q}}$ & $e^{-2i\rho}\vec{\tilde{t}}_{-\vec{k}}\cdot\vec{\tilde{t}}_{-\vec{k}+\vec{Q}}$ & $-\vec{\tilde{t}}_{-\vec{k}}\cdot\vec{\tilde{t}}_{-\vec{k}+\vec{Q}}$ & $-\vec{\tilde{t}}_{C\vec{k}}\cdot\vec{\tilde{t}}_{C\vec{k}+\vec{Q}}$ & $-\vec{\tilde{t}}_{-\vec{k}}\cdot\vec{\tilde{t}}_{-\vec{k}+\vec{Q}}$\tabularnewline
\hline 
$\vec{\tilde{t}}_{\vec{k}}\times\vec{\tilde{t}}_{\vec{k}+\vec{Q}}$ & $-\vec{\tilde{t}}_{C\vec{k}}\times\vec{\tilde{t}}_{C\vec{k}+\vec{Q}}$ & $-e^{-2i\rho}\vec{\tilde{t}}_{-\vec{k}}\times\vec{\tilde{t}}_{-\vec{k}+\vec{Q}}$ & $-e^{-2i\rho}\vec{\tilde{t}}_{-\vec{k}}\times\vec{\tilde{t}}_{-\vec{k}+\vec{Q}}$ & $-\vec{\tilde{t}}_{-\vec{k}}\times\vec{\tilde{t}}_{-\vec{k}+\vec{Q}}$ & $\vec{\tilde{t}}_{C\vec{k}}\times\vec{\tilde{t}}_{C\vec{k}+\vec{Q}}$ & $\vec{\tilde{t}}_{-\vec{k}}\times\vec{\tilde{t}}_{-\vec{k}+\vec{Q}}$\tabularnewline
\hline 
$\varepsilon_{\vec{k},z}$ & -- & -- & -- & $t_{-\vec{k},z}$ & $-t_{C\vec{k},z}$ & $t_{-\vec{k},z}$\tabularnewline
\hline 
\end{tabular}\caption{\label{Apptab:t_hybrid}The symmetry transformation rules under three
types of symmetries for $\vec{Q}$ in the symmetric gauge. $e^{-2i\rho}=e^{-2i\vec{Q}\cdot(\vec{\tilde{r}}_{B}-\vec{\tilde{r}}_{A})}=\pm1$.
$\varepsilon_{\vec{k},z}$ is not present in the type-1 case.}
\end{table*}
\end{widetext}

\section{Existence of a two-fold rotation symmetry $\{R|\vec{\tau}_{R}\}$
with $e^{2i\vec{Q}\cdot\vec{\tau}_{R}}=1$}

\label{App:existence_twofold_sym}In the main text, we propose a symmetry-related condition for the
emergence of an itinerant odd-parity antiferromagnetic phase in the
type-1 Wyckoff position system. This proposal requires a two-fold
site-preserving symmetry $\{R|\vec{\tau}_{R}\}$ satisfying $e^{2i\vec{Q}\cdot\vec{\tau}_{R}}=1$
for a given $\vec{Q}$. If such a symmetry is present, it is easy
to show that $\vec{t}_{\vec{k}}\times\vec{t}_{\vec{k}+\vec{Q}}=0$
on momentum $\vec{k}$ which is equivalent to $-R\vec{k}$ up to a
reciprocal lattice. Though both reflection (or glide) and rotation
(or screw) symmetries can serve this role, the cases with bands nesting
on a general point in a high-symmetry plane in the momentum space
is compatible with rotational symmetries. In this section, we discuss
when we have such two-fold rotation symmetries with $e^{2i\vec{Q}\cdot\vec{\tau}_{R}}=1$.

Regarding the type-1 Wyckoff position, let us recall the transformation
rules of $\vec{S}_{A}$ and $\vec{S}_{B}$ given in Eq.~\eqref{eq:symrule_type1} are derived from the transformation rule of the basis functions $f_{A}$ and $f_{B}$, which are written as

\begin{widetext}
\begin{equation}
\{{\cal I}|0\}\left(\begin{matrix}f_{A}\\
f_{B}
\end{matrix}\right)=\left(\begin{matrix} & 1\\
1
\end{matrix}\right)\left(\begin{matrix}f_{A}\\
f_{B}
\end{matrix}\right),\quad\{C|\vec{\tau}\}\left(\begin{matrix}f_{A}\\
f_{B}
\end{matrix}\right)=\left(\begin{matrix}1\\
 & -1
\end{matrix}\right)\left(\begin{matrix}f_{A}\\
f_{B}
\end{matrix}\right).\label{eq:symrule_basisfunctions}
\end{equation}
\end{widetext} In addition to these transformation rules, we can
add the transformation rule under a site-preserving symmetry $\{R|\vec{\tau}_{R}\}$
leaving $\vec{Q}$ invariant up to a reciprocal lattice vector:
\begin{equation}
\{R|\vec{\tau}_{R}\}\left(\begin{matrix}\vec{S}_{A}\\
\vec{S}_{B}
\end{matrix}\right)=\left(\begin{matrix}1\\
 & e^{2i\vec{Q}\cdot\vec{\tau}_{R}}
\end{matrix}\right)\left(\begin{matrix}\vec{S}_{A}\\
\vec{S}_{B}
\end{matrix}\right).\label{eq:symrule_basisfunctions2}
\end{equation}

Note that any properties of the space group should be compatible
with what can be drawn from these three transformation
rules in Eq.\textbf{~}\eqref{eq:symrule_basisfunctions} and Eq.\textbf{~}\eqref{eq:symrule_basisfunctions2}.
In particular, these tell that the product $f_{A}f_{B}$
is a basis for an one-dimensional even-parity irrep of the point group
of the space group. This implies that it should also be the basis for a one-dimensional
irreducible representation of a noncentrosymmetric point group which
is isomorphic to the site-symmetry group of an orbit of a Wyckoff position. Furthermore, $f_{A}f_{B}$ is even under $\{R|\vec{\tau}_{R}\}$
if and only if $e^{2i\vec{Q}\cdot\vec{\tau}_{R}}=1$. This let us examine the availability of a two-fold rotation $\{R|\vec{\tau}_{R}\}$
satisfying $e^{2i\vec{Q}\cdot\vec{\tau}_{R}}=1$ by looking into the irreducible representations of possible noncentrosymmetric subgroups of centrosymmetric point groups. To be specific, if  each one-dimensional irreducible representation of appropriate subgroups has at least one two-fold rotation represented by $+1$, it guarantees the availability of desired $\{R|\vec{\tau}_{R}\}$ in the corresponding space group.

For space groups $G$ of order $|G|$ including four-fold rotations,
all the one-dimensional irreps of the point groups of order $|G|/2$
should contain some four-fold rotations. This implies $f_{A}f_{B}$
is even under a two-fold rotation, which is equivalent to applying a four-fold rotation twice, as long as $\vec{Q}$ is invariant under the
four-fold rotation up to a reciprocal lattice vector. Consequently,
the existence of a site-preserving two-fold rotation $\{R|\vec{\tau}_{R}\}$
satisfying $e^{2i\vec{Q}\cdot\vec{\tau}_{R}}=1$ always exists in
such space groups. In particular, $C_{2z}$ serves as $R$ in primitive tetragonal lattices.
In consistency with this analysis, one can see that it is either $\{C_{2z}|\vec{\tau}\}$
with $\tau=0$ or $\tau=(1/2,1/2,0)$ that appears in the centrosymmetric
nonsymmorphic primitive tetragonal space groups so that $2\vec{Q}\cdot\vec{\tau}_{R}\in2\pi\mathbb{Z}$
for $\vec{Q}$ that is invariant under $C_{4z}$. The same applies to primitive
cubic systems where any of three two-fold rotations along the $x$,
$y$, and $z$ axes can serve as $R$. 

As for the primitive orthorhombic structure, only fifteen $D_{2h}$
symmetry structures are centrosymmetric and nonsymmorphic. Type-1 WPs are allowed only in five of them. Two point subgroups $C_{2v}$ and $D_{2d}$
of $D_{2h}$ point group can be isomorphic to the site-symmetry group
of type-1 Wyckoff position of multiplicity 2. As the subgroup $D_{2d}$
contains three two-fold rotations, at least one of them is represented
by $+1$ in all the one-dimensional irreps of $D_{2d}$. All the type-1
WPs in the space groups No. 48, 49, and 50 fall into this case. The
remaining two space groups are the space group No. 51 and No. 59, where
the site-symmetry group of a type-1 WP is isomorphic to $C_{2v}$
with the two-fold rotation $C_{2z}$ along the $z$-axis. In the space
group No. 51, $C_{2z}$ is followed by a half-translation $\tau_{R}=(1/2,0,0)$.
Therefore, $e^{2i\vec{Q}\cdot\vec{\tau}_{R}}=1$ is satisfied by all
nontrivial time-reversal invariant momentum $\vec{Q}=(0,y,z)$ with
$y,z=0,\pi$. In the space group No. 59, $C_{2z}$ is followed by
a half-translation $\tau_{R}=(1/2,1/2,0)$. Therefore, $\vec{Q}\in\{(\pi,\pi,0),(\pi,\pi,\pi),(0,0,\pi)\}$
satisfies $e^{2i\vec{Q}\cdot\vec{\tau}_{R}}=1$.

In the monoclinic, trigonal, and hexagonal lattices, we find all the
relevant point groups involves irreducible representations where all
two fold rotations are represented by $-1$ besides the trivial representation
with all symmetries represented by just $+1$. Thus, whether $\{R|\vec{\tau}_{R}\}$
satisfying $e^{2i\vec{Q}\cdot\vec{\tau}_{R}}=1$ is present or not
is not ensured in these systems.

\begin{widetext}
\section{Expression for $\beta_{i}$}

\label{App:beta}
A minimal microscopic Hamiltonian with two sublattice degrees of freedom
and Hubbard interaction of strength $U$ can be written as
\begin{align}
\hat{H} & =\sum_{\vec{k}}\hat{C}_{\vec{k}}^{\dagger}h_{\vec{k}}\hat{C}_{\vec{k}}+U\sum_{i,\tau}\hat{C}_{\vec{k},\tau,\uparrow}^{\dagger}\hat{C}_{\vec{k},\tau,\uparrow}\hat{C}_{\vec{k},\tau,\downarrow}^{\dagger}\hat{C}_{\vec{k},\tau,\downarrow}
\end{align}
with $\hat{C}_{\vec{k}}=(\hat{C}_{\vec{k},A,\uparrow},\hat{C}_{\vec{k},A,\downarrow},\hat{C}_{\vec{k},B,\uparrow},\hat{C}_{\vec{k},B,\downarrow})^{T}$ a spinor of annihilation operators, and $h_{\vec{k}}$ is defined in Eq.~\eqref{eq:hk} of the main text. The Hubbard interaction is decoupled via the Hubbard-Stratonovich transformation by introducing the auxiliary bosonic fields $\vec{S}_{A}$ and $\vec{S}_{B}$ which correspond to the magnetic order parameters with ordering vector
$\vec{Q}$. After integrating out the fermionic fields, we obtain the free energy
\begin{equation}
   {\cal F}=\sum_{\vec{k}\in{\rm FBZ}}{\rm Tr}\left\{\ln\left[\begin{pmatrix}
       1 & {\cal G}_{\vec{k}}\Sigma \\ {\cal G}_{\vec{k}+\vec{Q}}\Sigma^\dagger & 1
   \end{pmatrix}\right]\right\}+\frac{|\vec{S}_{A}|^{2}+|\vec{S}_{B}|^{2}}{U} 
\end{equation}
where ${\cal G}_{k} = (i\omega_n-h_{k})^{-1}$, while the precise form of $\Sigma$ depends on the gauge choice. The free energy up to quartic order in $\vec{S}_{A}$ and $\vec{S}_{B}$
is given by
\begin{align}
{\cal F}= & \alpha_{A}|\vec{S}_{A}|^{2}+\alpha_{B}|\vec{S}_{B}|^{2}+\alpha_{2}\vec{S}_{A}\cdot\vec{S}_{B}+\beta_{1}(|\vec{S}_{A}|^{2}+|\vec{S}_{B}|^{2})^{2}+\beta_{2}(\vec{S}_{A}\cdot\vec{S}_{B})^{2}+\beta_{3}|\vec{S}_{A}|^{2}|\vec{S}_{B}|^{2}\nonumber \\
 & +\beta_{4}(|\vec{S}_{A}|^{4}-|\vec{S}_{B}|^{4})+\{\beta_{5}(|\vec{S}_{A}|^{2}+|\vec{S}_{B}|^{2})+\beta_{6}(|\vec{S}_{A}|^{2}-|\vec{S}_{B}|^{2})\}\vec{S}_{A}\cdot\vec{S}_{B}.
\end{align}
Explicit expressions for the coefficients of the free energy Eq.~\eqref{eq:generalF} in the main text are derived by expanding the first
term up to quartic order in $\vec{S}_{A}$ and $\vec{S}_{B}$. Here,
we present the expressions in both conventional gauge and the symmetric
gauge introduced in Appendix~\ref{App:sym_t}.

\subsection{In the conventional gauge}

In the conventional gauge where the electronic Hamiltonian is periodic
in the momentum space, $h_{\vec{k}}$ and $\Sigma$ are given
by 
\begin{align}
h_{\vec{k}}= & \left(\begin{matrix}\epsilon_{\vec{k},0}+\epsilon_{\vec{k},z}-\mu & t_{\vec{k},x}-it_{\vec{k},y}\\
t_{\vec{k},x}+it_{\vec{k},y} & \epsilon_{\vec{k},0}-\epsilon_{\vec{k},z}-\mu
\end{matrix}\right),\quad\Sigma=\left(\begin{matrix} \vec{S}_{A}\cdot\vec{\sigma} & 0\\
 0 & \vec{S}_{B}\cdot\vec{\sigma}
\end{matrix}\right).
\end{align}

The quadratic coefficients are given by
\begin{align*}
\chi= & 2T\sum_{n,\vec{k}}g_{k}g_{k+Q}\{(i\omega_{n}-\varepsilon_{\vec{k},0})(i\omega_{n}-\varepsilon_{\vec{k}+\vec{Q},0})+t_{\vec{k},z}t_{\vec{k}+\vec{Q},z}\}\\
= & \sum_{\lambda,\lambda'=\pm}\sum_{\vec{k}}\frac{\tanh\frac{\xi_{\vec{k},\lambda}}{2T}-\tanh\frac{\xi_{\vec{k}+\vec{Q},\lambda'}}{2T}}{\xi_{\vec{k},\lambda}-\xi_{\vec{k}+\vec{Q},\lambda'}},\\
\alpha_{2}= & 4T\sum_{n,\vec{k}}\frac{\vec{t}_{\vec{k}}\cdot\vec{t}_{\vec{k}+\vec{Q}}}{\prod_{\lambda=\pm}(i\omega_{n}-\xi_{\vec{k},\lambda})(i\omega_{n}-\xi_{\vec{k}+\vec{Q},\lambda})},\\
\alpha_{3}= & 4T\sum_{n,\vec{k}}\frac{\varepsilon_{\vec{k}+\vec{Q},z}(i\omega_{n}-\varepsilon_{\vec{k},0})}{\prod_{\lambda=\pm}(i\omega_{n}-\xi_{\vec{k},\lambda})(i\omega_{n}-\xi_{\vec{k}+\vec{Q},\lambda})},
\end{align*}
with $\xi_{\vec{k},\pm}=\varepsilon_{\vec{k},0}\pm\sqrt{t_{\vec{k},x}^{2}+t_{\vec{k},y}^{2}+\varepsilon_{\vec{k},z}^{2}}$.
As summarized in Table~\ref{Apptab:t_conventional}, $\vec{t}_{\vec{k}}\cdot\vec{t}_{\vec{k}+\vec{Q}}$
is odd under the symmetry $\{C|\vec{\tau}\}$ satisfying $e^{2i\vec{Q}\cdot\vec{\tau}}$,
and thus $\alpha_{2}=0$. $\alpha_{3}$ vanishes because of a sublattice
transposing symmetry under which $\varepsilon_{\vec{k},z}$ is odd.

The quartic coefficients are 

\begin{align}
\beta_{1}= & T\sum_{k}\frac{\{(i\omega_{n}-\varepsilon_{\vec{k},0})^{2}+\varepsilon_{\vec{k},z}^{2}\}\{(i\omega_{n}-\varepsilon_{\vec{k}+\vec{Q},0})^{2}+\varepsilon_{\vec{k}+\vec{Q},z}^{2}\}}{\prod_{\lambda=\pm}\{(i\omega_{n}-\xi_{\vec{k},\lambda})(i\omega_{n}-\xi_{\vec{k}+\vec{Q},\lambda})\}^{2}}\nonumber \\
 & +4T\sum_{k}\frac{t_{\vec{k},z}t_{\vec{k}+\vec{Q},z}(i\omega_{n}-\varepsilon_{\vec{k}+\vec{Q},0})(i\omega_{n}-\varepsilon_{\vec{k},0})}{\prod_{\lambda=\pm}\{(i\omega_{n}-\xi_{\vec{k},\lambda})(i\omega_{n}-\xi_{\vec{k}+\vec{Q},\lambda})\}^{2}},\label{eq:beta1_conventional}\\
\beta_{2}= & 4T\sum_{i\omega,\vec{k}}\sum_{k}\frac{|\vec{t}_{\vec{k}}\cdot\vec{t}_{\vec{k}+\vec{Q}}|^{2}-|\vec{t}_{\vec{k}}\times\vec{t}_{\vec{k}+\vec{Q}}|^{2}}{\prod_{\lambda=\pm}\{(i\omega_{n}-\xi_{\vec{k},\lambda})(i\omega_{n}-\xi_{\vec{k}+\vec{Q},\lambda})\}^{2}},\label{eq:beta2_conventional}\\
\beta_{3}+2\beta_{1}= & 2T\sum_{i\omega,\vec{k}}\sum_{k}\frac{\{(i\omega_{n}-\varepsilon_{\vec{k},0})^{2}-\varepsilon_{\vec{k},z}^{2}\}|\vec{t}_{\vec{k}+\vec{Q}}|^{2}}{\prod_{\lambda=\pm}\{(i\omega_{n}-\xi_{\vec{k},\lambda})(i\omega_{n}-\xi_{\vec{k}+\vec{Q},\lambda})\}^{2}}\nonumber \\
 & +2T\sum_{i\omega,\vec{k}}\sum_{k}\frac{\{(i\omega_{n}-\varepsilon_{\vec{k}+\vec{Q},0})^{2}-\varepsilon_{\vec{k}+\vec{Q},z}^{2}\}|\vec{t}_{\vec{k}}|^{2}}{\prod_{\lambda=\pm}\{(i\omega_{n}-\xi_{\vec{k},\lambda})(i\omega_{n}-\xi_{\vec{k}+\vec{Q},\lambda})\}^{2}}\nonumber \\
 & +2T\sum_{i\omega,\vec{k}}\sum_{k}\frac{2(\vec{t}_{\vec{k}}\times\vec{t}_{\vec{k}+\vec{Q}})^{2}-|\vec{t}_{\vec{k}}|^{2}|\vec{t}_{\vec{k}+\vec{Q}}|^{2}}{\prod_{\lambda=\pm}\{(i\omega_{n}-\xi_{\vec{k},\lambda})(i\omega_{n}-\xi_{\vec{k}+\vec{Q},\lambda})\}^{2}},\label{eq:beta3_conventional}
\end{align}
and 
\begin{align}
\beta_{4}= & -2T\sum_{k}\frac{\{(i\omega_{n}-\varepsilon_{\vec{k},0})(i\omega_{n}-\varepsilon_{\vec{k}+\vec{Q},0})+\varepsilon_{\vec{k},z}\varepsilon_{\vec{k}+\vec{Q},z}\}}{\prod_{\lambda=\pm}\{(i\omega_{n}-\xi_{\vec{k},\lambda})(i\omega_{n}-\xi_{\vec{k}+\vec{Q},\lambda})\}^{2}}\nonumber \\
 & \times\{(i\omega_{n}-\varepsilon_{\vec{k},0})\varepsilon_{\vec{k}+\vec{Q},z}+(i\omega_{n}-\varepsilon_{\vec{k}+\vec{Q},0})\varepsilon_{\vec{k},z}\},\label{eq:beta4_conventional}\\
\beta_{5}= & 4T\sum_{k}\frac{\vec{t}_{\vec{k}}\cdot\vec{t}_{\vec{k}+\vec{Q}}\{(i\omega_{n}-\varepsilon_{\vec{k},0})(i\omega_{n}-\varepsilon_{\vec{k}+\vec{Q},0})+\varepsilon_{\vec{k},z}\varepsilon_{\vec{k}+\vec{Q},z}\}}{\prod_{\lambda=\pm}\{(i\omega_{n}-\xi_{\vec{k},\lambda})(i\omega_{n}-\xi_{\vec{k}+\vec{Q},\lambda})\}^{2}},\label{eq:beta5_conventional}\\
\beta_{6}= & 4T\sum_{k}\frac{\vec{t}_{\vec{k}}\cdot\vec{t}_{\vec{k}+\vec{Q}}\{\varepsilon_{\vec{k}+\vec{Q},z}(i\omega_{n}-\varepsilon_{\vec{k},0})+\varepsilon_{\vec{k},z}(i\omega_{n}-\varepsilon_{\vec{k}+\vec{Q},0})\}}{\prod_{\lambda=\pm}\{(i\omega_{n}-\xi_{\vec{k},\lambda})(i\omega_{n}-\xi_{\vec{k}+\vec{Q},\lambda})\}^{2}}.\label{eq:beta6_conventional}
\end{align}
where it can be easily checked that $\beta_{5}=\beta_{6}=0$ due to
$\{C|\vec{\tau}\}$ and $\beta_{4}=0$ because $\varepsilon_{\vec{k},z}$
is odd under a sublattice transposing symmetry.

Important combinations of $\beta_{1}$, $\beta_{2}$, and $\beta_{3}$
are 

\begin{align}
\beta_{3}= & 2\gamma+2T\sum_{n,\vec{k}}\frac{|\vec{t}_{\vec{k}}\times\vec{t}_{\vec{k}+\vec{Q}}|^{2}}{\prod_{\lambda=\pm}\{(i\omega_{n}-\xi_{\vec{k},\lambda})(i\omega_{n}-\xi_{\vec{k}+\vec{Q},\lambda})\}^{2}},\label{eq:boundary_btw_1_and_3}\\
\beta_{2}+\beta_{3}= & 2\gamma+2T\sum_{n,\vec{k}}\frac{|\vec{t}_{\vec{k}}\cdot\vec{t}_{\vec{k}+\vec{Q}}|^{2}}{\prod_{\lambda=\pm}\{(i\omega_{n}-\xi_{\vec{k},\lambda})(i\omega_{n}-\xi_{\vec{k}+\vec{Q},\lambda})\}^{2}},\label{eq:boundary_btw_1_and_2}
\end{align}
with 
\begin{equation}
\gamma=T\sum_{i\omega_{n},k}\frac{1}{\prod_{\lambda=\pm}(i\omega_{n}-\xi_{\vec{k},\lambda})(i\omega_{n}-\xi_{\vec{k}+\vec{Q},\lambda})}=-\frac{1}{8}\sum_{\vec{k}}\frac{1}{|\vec{t}_{\vec{k}}||\vec{t}_{\vec{k}+\vec{Q}}|}\sum_{\lambda,\lambda'=\pm}\lambda\lambda'\frac{\tanh\frac{\xi_{\vec{k},\lambda}}{2T}-\tanh\frac{\xi_{\vec{k}+\vec{Q},\lambda'}}{2T}}{\xi_{\vec{k},\lambda}-\xi_{\vec{k}+\vec{Q},\lambda'}}.
\end{equation}

When the inter-band nesting (: nesting between $\xi_{\vec{k},\lambda}$
and $\xi_{\vec{k}+\vec{Q},-\lambda}$) occurs, $\gamma>0$ is expected
near the peak of $\chi$ if $|\vec{t}_{\vec{k}}\times\vec{t}_{\vec{k}+\vec{Q}}|^{2}$
is vanisingly small around the region where the nesting between the
bands occurs. Therefore, the OP-AFM phase will be favored in this
case. If the instability is driven by an intra-band nesting (: nesting
between $\xi_{\vec{k},\lambda}$ and $\xi_{\vec{k}+\vec{Q},\lambda}$)
, this would result in the spin-charge phase. 

\subsection{In the symmetric gauge}

In the symmetric gauge where the representation matrices of symmetry operations are independent of momentum, $h_{\vec{k}}$ and $\Sigma$ are given
by 
\begin{align}
h_{\vec{k}}= & \left(\begin{matrix}\epsilon_{\vec{k},0}+\epsilon_{\vec{k},z}-\mu & \tilde{t}_{\vec{k},x}-i\tilde{t}_{\vec{k},y}\\
\tilde{t}_{\vec{k},x}+i\tilde{t}_{\vec{k},y} & \epsilon_{\vec{k},0}-\epsilon_{\vec{k},z}-\mu
\end{matrix}\right),\quad\Sigma=\left(\begin{matrix} \vec{S}_{A}\cdot\vec{\sigma} & 0\\
 0 & e^{i\rho}\vec{S}_{B}\cdot\vec{\sigma}
\end{matrix}\right).
\end{align}
with $\rho=\vec{Q}\cdot(\vec{r}_{B}-\vec{r}_{A})$ and $\tilde{t}_{\vec{k},x}-i\tilde{t}_{\vec{k},y}$ is related with
$t_{\vec{k},x}-it_{\vec{k},y}$ through $\tilde{t}_{\vec{k},x}-i\tilde{t}_{\vec{k},y}=(t_{\vec{k},x}-it_{\vec{k},x})e^{i\vec{k}(\vec{r}_{A}-\vec{r}_{B})}$.
Moreover, 
\begin{align*}
\vec{\tilde{t}}_{\vec{k}}\cdot\vec{\tilde{t}}_{\vec{k}+\vec{Q}}= & {\rm Re}[(t_{\vec{k},x}-it_{\vec{k},y})(t_{\vec{k}+\vec{Q},x}+it_{\vec{k}+\vec{Q},x})e^{i\rho}]\\
= & \cos\rho\cdot(\vec{t}_{\vec{k}}\cdot\vec{t}_{\vec{k}+\vec{Q}})-\sin\rho(\vec{t}_{\vec{k}}\times\vec{t}_{\vec{k}+\vec{Q}})_{z},\\
(\vec{\tilde{t}}_{\vec{k}}\times\vec{\tilde{t}}_{\vec{k}+\vec{Q}})_{z}= & {\rm Im}[(t_{\vec{k},x}-it_{\vec{k},y})(t_{\vec{k}+\vec{Q},x}+it_{\vec{k}+\vec{Q},x})e^{i\rho}]\\
= & \sin\rho\cdot(\vec{t}_{\vec{k}}\cdot\vec{t}_{\vec{k}+\vec{Q}})+\cos\rho(\vec{t}_{\vec{k}}\times\vec{t}_{\vec{k}+\vec{Q}})_{z}.
\end{align*}
One should note that $e^{i\rho}=e^{i\vec{Q}\cdot(\vec{\tilde{r}}_{B}-\vec{\tilde{r}}_{A})}=\pm i,\pm1$
for the type-1 case, while $e^{i\rho}=e^{i\vec{Q}\cdot\vec{\tau}}=\pm i$
for the type-2 case. The former result is drawn from the observation
that in all centrosymmetric nonsymmorphic tetragonal or orthorombic
space groups, each flattened coordinate $a$, $b$, $c$ of $\vec{\tilde{r}}=(a,b,c)$
of sublattices of a type-1 WP of multiplicty 2 belongs to the set
$\{0,1/4,1/2,3/4\}$.

\begin{align}
\beta_{2}= & 4T\cos2\rho\sum_{\omega_{n},\vec{k}}\frac{|\vec{\tilde{t}}_{\vec{k}}\cdot\vec{\tilde{t}}_{\vec{k}+\vec{Q}}|^{2}-|\vec{\tilde{t}}_{\vec{k}}\times\vec{\tilde{t}}_{\vec{k}+\vec{Q}}|^{2}}{\prod_{\lambda=\pm}\{(i\omega_{n}-\xi_{\vec{k},\lambda})(i\omega_{n}-\xi_{\vec{k}+\vec{Q},\lambda})\}^{2}}\label{eq:beta2_Hybrid}\\
 & +4T\sin2\rho\sum_{\omega_{n},\vec{k}}\frac{2(\vec{\tilde{t}}_{\vec{k}}\cdot\vec{\tilde{t}}_{\vec{k}+\vec{Q}})(\vec{\tilde{t}}_{\vec{k}}\times\vec{\tilde{t}}_{\vec{k}+\vec{Q}})}{\prod_{\lambda=\pm}\{(i\omega_{n}-\xi_{\vec{k},\lambda})(i\omega_{n}-\xi_{\vec{k}+\vec{Q},\lambda})\}^{2}}\nonumber \\
\beta_{5}= & 4T\sum_{k}\frac{\vec{\tilde{t}}_{\vec{k}}\cdot\vec{\tilde{t}}_{\vec{k}+\vec{Q}}\cos\rho+(\vec{\tilde{t}}_{\vec{k}}\times\vec{\tilde{t}}_{\vec{k}+\vec{Q}})_{z}\sin\rho}{\prod_{\lambda=\pm}\{(i\omega_{n}-\xi_{\vec{k},\lambda})(i\omega_{n}-\xi_{\vec{k}+\vec{Q},\lambda})\}^{2}}\\
 & \times\{(i\omega_{n}-\varepsilon_{\vec{k},0})(i\omega_{n}-\varepsilon_{\vec{k}+\vec{Q},0})+\varepsilon_{\vec{k},z}\varepsilon_{\vec{k}+\vec{Q},z}\},\nonumber \\
\beta_{6}= & 4T\sum_{k}\frac{\vec{\tilde{t}}_{\vec{k}}\cdot\vec{\tilde{t}}_{\vec{k}+\vec{Q}}\cos\rho+(\vec{\tilde{t}}_{\vec{k}}\times\vec{\tilde{t}}_{\vec{k}+\vec{Q}})_{z}\sin\rho}{\prod_{\lambda=\pm}\{(i\omega_{n}-\xi_{\vec{k},\lambda})(i\omega_{n}-\xi_{\vec{k}+\vec{Q},\lambda})\}^{2}}\\
 & \times\{\varepsilon_{\vec{k}+\vec{Q},z}(i\omega_{n}-\varepsilon_{\vec{k},0})+\varepsilon_{\vec{k},z}(i\omega_{n}-\varepsilon_{\vec{k}+\vec{Q},0})\},\nonumber 
\end{align}
while $\beta_{1}$, $\beta_{3}$ and $\beta_{4}$, whose numerators
do not involving $\vec{t}_{\vec{k}}\cdot\vec{t}_{\vec{k}+\vec{Q}}$
and $\vec{t}_{\vec{k}}\times\vec{t}_{\vec{k}+\vec{Q}}$ in Eqs.~\eqref{eq:beta1_conventional},
~\eqref{eq:beta3_conventional}, and~\eqref{eq:beta4_conventional},
are the same as those give in Eqs.~\eqref{eq:beta1_conventional},
~\eqref{eq:beta3_conventional}, and~\eqref{eq:beta4_conventional}.

Note that $\cos2\rho=\pm1$ and $\sin2\rho=0$, and thus the second
term of Eq.~\eqref{eq:beta2_Hybrid} vanishes. Furthermore, the $\vec{\tilde{t}}_{\vec{k}}\cdot\vec{\tilde{t}}_{\vec{k}+\vec{Q}}$
and $(\vec{\tilde{t}}_{\vec{k}}\times\vec{\tilde{t}}_{\vec{k}+\vec{Q}})_{z}$
are odd under the sublattice-preserving symmetry which makes $\beta_{5}$
and $\beta_{6}$ null.

\section{Spin-spliting of the bands}

\label{App:SpinSpliting}Here, we provide the condition for lifting
the degeneracy of the bands from the (conventional-gauge) Hamiltonian matrix
\begin{align}
H_{\vec{k}} & =\left(\begin{matrix}h_{\vec{k}} & \Sigma\\
\Sigma & h_{\vec{k}+\vec{Q}}
\end{matrix}\right),\label{eq:Hk}
\end{align}
where $h_{\vec{k}}=[\varepsilon_{\vec{k},0}\tau_{0}+t_{\vec{k},x}\tau_{x}+t_{\vec{k},y}\tau_{y}+\varepsilon_{\vec{k},z}\tau_{z}]\sigma_{0}$
and $\Sigma=(\vec{S}_{A}\cdot\vec{\sigma})(\tau_{0}+\tau_{z})/2 +(\vec{S}_{B}\cdot\vec{\sigma})(\tau_{0}-\tau_{z})/2$. We evaluate the characteristic polynomial $\det(\lambda-H_{\vec{k}})$
of $H_{\vec{k}}$ for the three stable saddle points:
\begin{enumerate}
\item For the half-colinear state with $(\vec{S}_{A},\vec{S}_{B})=S(\hat{x},0)$,
\begin{align}
\det(\lambda-H_{\vec{k}})= \big[\{(\varepsilon_{\vec{k},0}-\lambda)^{2}-\delta\varepsilon_{\vec{k}}^{2}\}\{(\varepsilon_{\vec{k}+\vec{Q},0}-\lambda)^{2}-\delta\varepsilon_{\vec{k}+\vec{Q}}^{2}\}-S^{2}(\lambda-\varepsilon_{\vec{k},0}+\varepsilon_{\vec{k},z})(\lambda-\varepsilon_{\vec{k}+\vec{Q},0}+\varepsilon_{\vec{k}+\vec{Q},z})\big]^{2}
\end{align}
with $\delta\varepsilon_{\vec{k}}=\sqrt{t_{\vec{k},x}^{2}+t_{\vec{k},y}^{2}+\varepsilon_{\vec{k},z}^{2}}$.
Note that every zero of the characteristic polynomial is doubly-degenerate.
\item For the colinear state with $(\vec{S}_{A},\vec{S}_{B})=S(\hat{x},\hat{x})$,
\begin{align}
\det(\lambda-H_{\vec{k}})= & \big[\{(\varepsilon_{\vec{k},0}-\lambda)^{2}-\delta\varepsilon_{\vec{k}}^{2}\}\{(\varepsilon_{\vec{k}+\vec{Q},0}-\lambda)^{2}-\delta\varepsilon_{\vec{k}+\vec{Q}}^{2}\}+S^{4}\\
 & -2S^{2}\{(\varepsilon_{\vec{k},0}-\lambda)(\varepsilon_{\vec{k}+\vec{Q},0}-\lambda)+\varepsilon_{\vec{k},z}\varepsilon_{\vec{k}+\vec{Q},z}-\vec{t}_{\vec{k}}\cdot\vec{t}_{\vec{k}+\vec{Q}}\}\big]^{2},\nonumber 
\end{align}
with $\vec{t}_{\vec{k}}\cdot\vec{t}_{\vec{k}+\vec{Q}}=t_{\vec{k},x}t_{\vec{k}+\vec{Q},x}+t_{\vec{k},y}t_{\vec{k}+\vec{Q},y}$.
Again, all the zeros of the characteristic polynomial are doubly-degenerate.
\item For the coplanar state with $(\vec{S}_{A},\vec{S}_{B})=S(\hat{x},\hat{y})$,
\begin{align}
\det(\lambda-H_{\vec{k}})= & \big[\{(\varepsilon_{\vec{k},0}-\lambda)^{2}-\delta\varepsilon_{\vec{k}}^{2}\}\{(\varepsilon_{\vec{k}+\vec{Q},0}-\lambda)^{2}-\delta\varepsilon_{\vec{k}+\vec{Q}}^{2}\}+S^{4}\\
 & -2S^{2}\{(\varepsilon_{\vec{k},0}-\lambda)(\varepsilon_{\vec{k}+\vec{Q},0}-\lambda)+\varepsilon_{\vec{k},z}\varepsilon_{\vec{k}+\vec{Q},z}-\vec{t}_{\vec{k}}\times\vec{t}_{\vec{k}+\vec{Q}}\}\big]\nonumber \\
 & \times\big[\{(\varepsilon_{\vec{k},0}-\lambda)^{2}-\delta\varepsilon_{\vec{k}}^{2}\}\{(\varepsilon_{\vec{k}+\vec{Q},0}-\lambda)^{2}-\delta\varepsilon_{\vec{k}+\vec{Q}}^{2}\}+S^{4}\nonumber \\
 & -2S^{2}\{(\varepsilon_{\vec{k},0}-\lambda)(\varepsilon_{\vec{k}+\vec{Q},0}-\lambda)+\varepsilon_{\vec{k},z}\varepsilon_{\vec{k}+\vec{Q},z}+\vec{t}_{\vec{k}}\times\vec{t}_{\vec{k}+\vec{Q}}\}\big]\nonumber
\end{align}
with $\vec{t}_{\vec{k}}\times\vec{t}_{\vec{k}+\vec{Q}}=t_{\vec{k},x}t_{\vec{k}+\vec{Q},y}-t_{\vec{k},y}t_{\vec{k}+\vec{Q},x}$.
Note that the zeros of the characteristic polynomial are nondegenerate
only when $\vec{t}_{\vec{k}}\times\vec{t}_{\vec{k}+\vec{Q}}\neq0$.
If $\vec{t}_{\vec{k}}\times\vec{t}_{\vec{k}+\vec{Q}}=0$, $H_{\vec{k}}$
in Eq. \eqref{eq:Hk} can be transformed into a symmetric matrix with
real components, and this reaility keeps the double-degeneracy of
the bands.
\end{enumerate}
In the calculation above we have chosen specific orientations of the staggered order parameters without loss of generality, since only the relative orientations are significant due to the spin-rotation symmetry.

An explict eigenvalues of $H_{\vec{k}}$ in Eq. \eqref{eq:Hk} can
be derived if we neglect the trivial term $\varepsilon_{\vec{k},0}-\mu$
which has nothing to do with the symmetry properties of the normal
phase. The eigenvalues of the $8\times8$ Hamiltonian are given by
\citep{Yue2025}
\begin{equation}
\xi_{\pm,\pm',\pm''}=\pm\frac{1}{2}\sqrt{X\pm'\sqrt{X^{2}-Y\pm''Z}},\label{eq:Dispersion}
\end{equation}
with 
\begin{align}
X= & \vec{S}_{A}^{2}+\vec{S}_{B}^{2}+\vec{t}_{\vec{k}}^{2}+\vec{t}_{\vec{k}+\vec{Q}}^{2},\\
Y= & 4(\vec{S}_{A}^{2}\vec{S}_{B}^{2}+\vec{t}_{\vec{k}}^{2}\vec{t}_{\vec{k}+\vec{Q}}^{2})-8(\vec{S}_{A}\cdot\vec{S}_{B})(\vec{t}_{\vec{k}}\cdot\vec{t}_{\vec{k}+\vec{Q}}),\\
Z= & 8|\vec{S}_{A}\times\vec{S}_{B}||\vec{t}_{\vec{k}}\times\vec{t}_{\vec{k}+\vec{Q}}|.
\end{align}
Eight distinct dispersions are obtained when $|\vec{S}_{A}\times\vec{S}_{B}||\vec{t}_{\vec{k}}\times\vec{t}_{\vec{k}+\vec{Q}}|\neq0$
in accordance with the analysis of the multiplicity of the zeros of
the characteristic polynomial $\det(\lambda-H_{\vec{k}})$.

\end{widetext}

\end{document}